\documentclass[a4paper,fleqn]{cas-dc}

\usepackage[numbers,sort,compress]{natbib}
\usepackage{subcaption}

\def\tsc#1{\csdef{#1}{\textsc{\lowercase{#1}}\xspace}}
\tsc{WGM}
\tsc{QE}
\begin{document}
\renewcommand{\printorcid}{}
\sloppy
\let\WriteBookmarks\relax
\def\floatpagepagefraction{1}
\def\textpagefraction{.001}

% Short title
\shorttitle{Time-Frequency Distributions of Heart Sound Signals}
% Short author
\shortauthors{X. Bao et~al.}  

% Main title of the paper
\title [mode = title]{Time-Frequency Distributions of Heart Sound Signals: A Comparative Study using Convolutional Neural Networks}  
% \tnotemark[1]
% \tnotetext[1]{This study was supported in part by the King’s-China Scholarship Council PhD Scholarship programme, and has been funded in part from KFAS, Kuwait Foundation for Advancement of Sciences, project no. CN20-13EE-01. This study used the Cirrus UK National Tier-2 HPC Service at EPCC (\url{http://www.cirrus.ac.uk}) funded by the University of Edinburgh and EPSRC (EP/P020267/1).}

\author[1]{Xinqi Bao}
\cormark[1]
\credit{Conceptualization, Methodology, Writing original draft}
\author[1]{Yujia Xu}
\cormark[1]
\credit{Methodology, Writing original draft}
\author[1]{Hak-Keung Lam}
\credit{Review \& editing, Supervision}
\author[2]{Mohamed Trabelsi}
\credit{Funding acquisition, Project administration, Review \& editing}
\author[3]{Ines Chihi}
\credit{Review \& editing}
\author[4]{Lilia Sidhom}
\credit{Review \& editing}
\author[1,5]{Ernest N. Kamavuako}
\cormark[2]
\credit{Conceptualization, Method improvement, Review \& editing, Supervision}
\ead{ernest.kamavuako@kcl.ac.uk}

\cortext[cor1]{These authors contributed equally to this work.}
\cortext[cor2]{Corresponding author}

% Address/affiliation
\affiliation[1]{organization={Department of Engineering},
                addressline={King's College London, Strand}, 
                city={London},
                postcode={WC2R 2LS},
                country={United Kingdom}}
\affiliation[2]{organization={Department of Electronic and Communications Engineering},
            addressline={Kuwait College of Science and Technology}, 
            % city={},
            % postcode={}, 
            % state={},
            country={Kuwait}}
\affiliation[3]{organization={Department of Engineering},
            addressline={Université du Luxembourg}, 
            % city={},
            postcode={1359},
            country={Luxembourg}}
\affiliation[4]{organization={National Engineering School of Bizerta},
            addressline={Carthage University}, 
            city={Tunis},
            postcode={2070},
            country={Tunisia}}
            
\affiliation[5]{organization={Faculté de Médecine},
            addressline={Université de Kindu}, 
            city={Kindu},
            % postcode={2070},
            country={Democratic Republic of the Congo}}
\begin{abstract}
Time-Frequency Distributions (TFDs) support the heart sound characterisation and classification in early cardiac screening. However, despite the frequent use of TFDs in signal analysis, no study comprehensively compared their performances on deep learning for automatic diagnosis. Furthermore, the combination of signal processing methods as inputs for Convolutional Neural Networks (CNNs) has been proved as a practical approach to increasing signal classification performance. Therefore, this study aimed to investigate the optimal use of TFD/ combined TFDs as input for CNNs. The presented results revealed that: 1) The transformation of the heart sound signal into the TF domain achieves higher classification performance than using of raw signals. Among the TFDs, the difference in the performance was slight for all the CNN models (within $1.3\%$ in average accuracy). However, Continuous wavelet transform (CWT) and Chirplet transform (CT) outperformed the rest. 2) The appropriate increase of the CNN capacity and architecture optimisation can improve the performance, while the network architecture should not be overly complicated. Based on the ResNet or SEResNet family results, the increase in the number of parameters and the depth of the structure do not improve the performance apparently. 3) Combining TFDs as CNN inputs did not significantly improve the classification results. The findings of this study provided the knowledge for selecting TFDs as CNN input and designing CNN architecture for heart sound classification.
\end{abstract}

\maketitle

\section{Introduction}
\label{sec:introduction}
Cardiovascular diseases (CVDs) refer to unhealthy conditions of the heart and related blood vessels \cite{mendis2011global}, such as congenital heart disease, coronary artery disease, valvular heart disease, arrhythmias, heart failure, etc. According to the World Health Organization (WHO), CVDs are the leading causes of death, with a mortality rate of $32\%$ of all global deaths \cite{who2021cvds}. Early screening and intervention can prevent CVDs from worsening into severe health issues. Auscultation is the most common and practical approach in early cardiac screening. The physician uses a stethoscope to listen to the heart sound of the patient and diagnose the heart conditions. However, it relies significantly on the medical staff's clinical experience and listening ability. According to existing surveys, experienced cardiologists can distinguish pathological murmurs with about $80\%$ accuracy, while inexperienced new physicians or trainees can distinguish them with less than $40\%$ accuracy \cite{kumar2013evaluation, lam2005factors}. On one side, misdiagnosis might occur and lead to serious health issues. On the other side, false-positive cases increase health costs because of further investigation, such as the electrocardiogram (ECG), cardiac ultrasound, computerised tomography, wasting medical resources, and reducing diagnostic efficiency for hospitals and medical staff. 

Computer-aided heart sound analysis is a potential method to improve auscultation accuracy. It can overcome the clinical experience and listening limitation by using machine learning algorithms to learn the diagnostic ability on large amounts of data. The first published study on automatic heart sound classification can be traced back to 1963 when research used threshold to identify rheumatic heart disease \cite{gerbarg1963computer}. Afterwards, more Machine Learning (ML) techniques have been explored, such as logistic regression \cite{bobillo2016tensor}, regression tree \cite{amiri2013intelligent}, K-nearest neighbours (KNN) \cite{bobillo2016tensor, safara2013multi, jaramillo2008feature}, random forest \cite{singh2016using}, support vector machine (SVM) \cite{bobillo2016tensor, liu2012autonomous, patidar2015automatic}, hidden Markov model (HMM) \cite{wang2007phonocardiographic}, etc. Besides traditional ML approaches, neural network (NN) and its variants have also been applied to heart sound classification \cite{dokur2008heart,phanphaisarn2011heart,ari2009search}. With the development of NN and advanced computing hardware, more dedicated and deeper NNs have been proposed, such as convolutional neural networks (CNNs). CNN focuses on image recognition; and recurrent neural network (RNN) is specialised in processing sequence data such as audio and text \cite{zeyer2017comprehensive, liu2019bidirectional}. Compared with traditional ML methods, deep learning algorithms can neglect manual feature extraction and use the raw signal as input with promising performance. After the $2016$ PhysioNet/Computing in Cardiology (CinC) Challenge \cite{liu2016open}, using CNN or RNN to conduct heart sound classification became the mainstream approach \cite{huai2020heart,rubin2017recognizing,demir2019towards,dominguez2017deep,cheng2019design,ergen2012time}.

\begin{figure}[!ht]
    \centering
    \begin{subfigure}{\columnwidth} \centering
        \includegraphics[width=0.65\columnwidth]{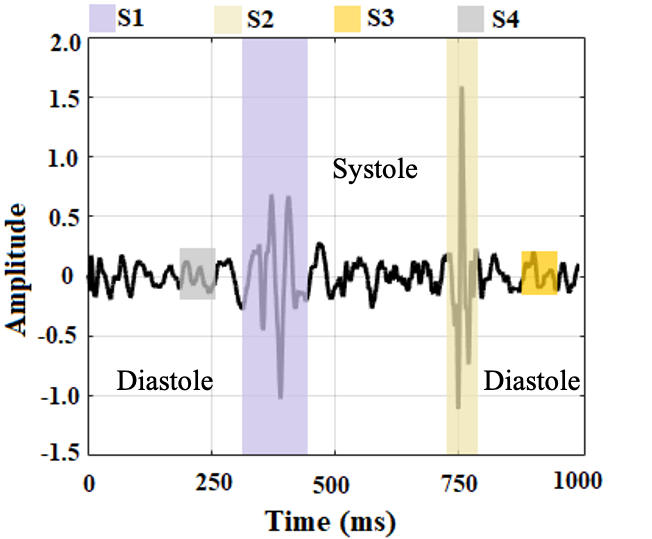}
        \caption{}
    \label{subfig:heart_sound_location}
    \end{subfigure}
    \begin{subfigure}{\columnwidth} \centering
        \includegraphics[width=0.65\columnwidth]{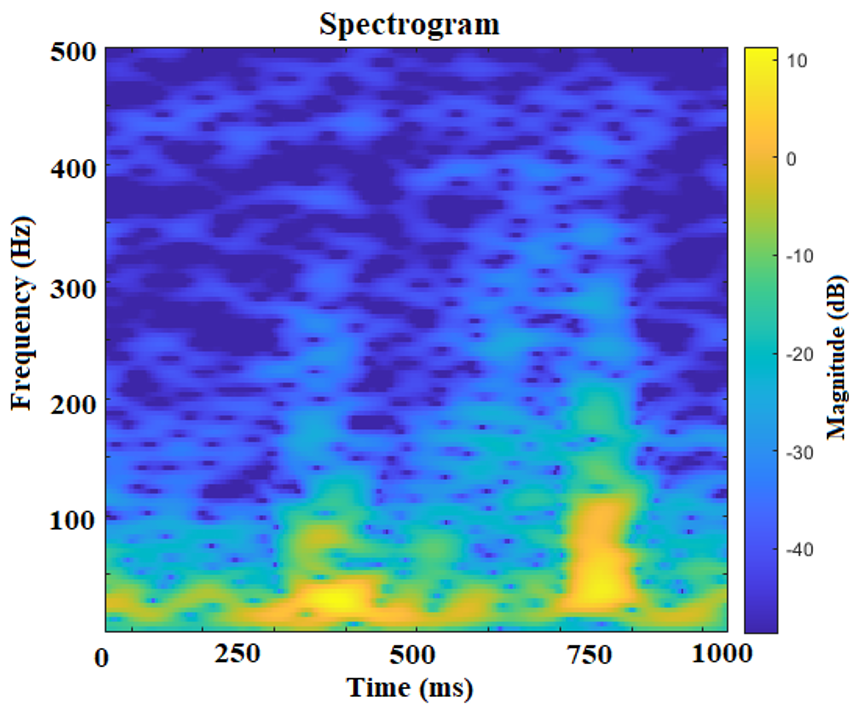}
        \caption{}
        \label{subfig:heart_sound_spectrogram}
    \end{subfigure}
    \caption{(a) Visualisation of heart sound signal with its component locations. (b) Spectrogram of the heart sound.}
    \label{fig:heart_sound_visualization}
\end{figure}

Phonocardiogram (PCG) is the waveform of the heart sound captured by the acoustic sensors constituting the raw signal for the heart sound analysis as shown in Fig.~\ref{subfig:heart_sound_location}. For healthy conditions, S1 and S2 are the two main components that represent the closure of mitral and tricuspid valves (S1) and the closure of aortic and pulmonic valves (S2), respectively. However, there could be murmurs in the waveform and according to their location and morphology, they can reflect different kinds of heart conditions. For instance, one can locate the aortic stenosis murmurs during the systole between S1 and S2. S3 and S4 are innocent components seen in children's PCG signals but barely in adults. Since the PCG signals are low in frequency (below $70$ Hz) and short in time (S1 approximate $120$ ms and S2 approximate $100$ ms), using time-domain representation only cannot reveal all the contained information. Therefore, in the heart sound analysis, the PCG signals are typically transformed into the time-frequency domain for better visualisation and processing, as shown in Fig.~\ref{subfig:heart_sound_spectrogram} for the case of the spectrogram.

Time-frequency distributions (TFDs) represent signals in the time and frequency domains, used in a wide range of signal analyses, such as optics, acoustics, and biomedicine \cite{stankovic2014time}. For bio-signals, TFDs are used to analyse electroencephalography (EEG), electromyography (EMG), ECG, etc \cite{mahmoud2006time}. TFDs transform a 1-D signal in the time domain into a 2-D space, containing both time and frequency information. This transformation can provide a more intuitive visualisation of the frequency information. Furthermore, as inputs for CNNs, they can also reduce the difficulty of frequency information mining. For PCG classification, the murmurs are defined in time and frequency, therefore, the availability of this information to the classifier is a reasonable basis for improvement.
% \textcolor{blue}{TFDs transform a 1-D signal in the time domain into a 2-D space, containing both time and frequency information. This transformation can provide a more intuitive visualisation of the frequency information. Furthermore, as inputs for CNNs, they can also reduce the difficulty of frequency information mining. For PCG classification, the murmurs are defined in time and frequency, therefore, the availability of this information to the classifier is a reasonable basis for improvement.}
The short-time Fourier transform (STFT) is the most commonly used transform, which is relatively simple to calculate. However, it cannot effectively track the abrupt changes because a trade-off is needed between the time and frequency windows’ size. This means that an increased time-domain resolution will worsen the resolution in the frequency domain and vice versa \cite{obaidat1993phonocardiogram,peng2011polynomial}. The continuous wavelet transform (CWT) is another approach for time-frequency analysis. Compared with STFT, it can capture more local information by scaling and shifting the mother wavelet to fit the raw signal. This improves the resolution but could not provide greater frequency resolution at higher frequency bands \cite{auger1996time}. However, due to the low-frequency characteristics of the heart sound, this drawback of CWT can be neglected. Thus, CWT is often used in PCG analysis \cite{ergen2012time,cherif2010choice,vikhe2009wavelet,debbal2004analysis}. The Chirplet transform (CT) can be regarded as a generalisation of both STFT and CWT, including four parameters: time, frequency, scale, and chirp rate \cite{peng2011polynomial,taebi2017analysis,mann1995chirplet}. It also holds the advantage of detecting instantaneous frequency trajectory used in PCG analysis  \cite{ghosh2020automated,taebi2017analysis}. In addition to the mentioned TFDs, Cohen's class distributions such as Wigner-Ville distribution (WVD) and Choi-Williams distribution (CWD) were also applied in the heart sound analysis \cite{djebbari2013detection,chen2015recognition}. They can provide higher resolution, but with cross-term issues (also known as interference terms). In \cite{taebi2017time}, several TFDs were graphically compared and analysed on their ability to visualise PCG events.
% \textcolor{blue}{In \cite{taebi2017time}, several TFDs were graphically compared and analysed on their ability to visualise PCG events.}

With the development of computer vision, deep CNN is partially undertaking the analysis task to provide the auxiliary diagnosis. The TFDs were used as inputs to train the deep learning algorithms \cite{huai2020heart,bozkurt2018study,ren2018learning,nilanon2016normal,rizalclassification,alaskar2019implementation,demir2019towards,dominguez2017deep,cheng2019design}. However, due to the different databases, inputs, and network architectures, it is unclear how selecting TFDs, and CNNs can affect heart sound classification. Furthermore, many studies have proved that combining different signal processing methods can improve classification performance \cite{jung2021efficiently,zhang2019large,mcloughlin2020time,jalayer2021fault,liu2021multiscale,liu2021ecg}. Though, their combining methods differ significantly, including channel-wise stacking \cite{jalayer2021fault}, spatial concatenation \cite{jung2021efficiently}, hidden feature fusing \cite{jalayer2021fault,liu2021multiscale,liu2021ecg} and input vector concatenation \cite{yan2018novel}. Among the combination methods, channel-wise stacking is the mainstream.

Nevertheless, this approach has received less attention in the heart sound classification field. Hence, it is necessary to investigate whether the channel-wise stacking method to combine the different TFDs can further improve the heart sound classification. 

Overall, the aims of this study are: 1) to assess if the selection of the TFDs will affect heart sound classification; 2) to compare the performance of several state-of-the-art CNNs with different capacity, depth, and architecture, for different TFDs; 3) to ascertain if combined TFDs instead of single TFD as input can improve classification accuracy. Thus, the main contribution of this study is to provide insight into the selection of CNN inputs (in terms of TFDs) when designing heart sound classification methods. 

\section{Methodology}
\label{sec:method}
In this study, five TFD approaches were encompassed and compared as CNN inputs, including STFT, CWT, CT, WVD, and CWD. The CNN models were selected with different complexity and computing load, including MobileNetV3-Small/Large, ResNet-18/34/50, SEResNet-18/34/50, DenseNet-121/169. This section provides the theories and properties of the TFD methods. This section also includes a description of the CNNs models and experimental settings.

\subsection{Time-frequency distribution Methods}
\label{subsec:TFD-methods}
1) Short-time Fourier transform \\
\indent Fourier transform is a linear integral transform that converts the signal $x(t)$ from the time domain into the frequency domain $\widehat{X}$. To provide simultaneous time and frequency information, STFT is applied with a sliding window function $\omega(t)$ on the signal conducting a fast Fourier transform (FFT) within the windows to determine the frequency variation over time. Eq. (\ref{eq:stft}) defines the STFT:
\begin{equation}
    % \hat{\mathbf{X}}_{STFT} = \lim
    \widehat{X}_{STFT}=\int_{-\infty}^{\infty} w(t-\tau) x(\tau) e^{-j \omega \tau} d \tau
\label{eq:stft}
\end{equation}

\noindent where the transform output $\widehat{X}_{STFT}$ can be regarded as a time-dependent frequency spectrum (spectrogram), $t$ and $\omega$ are the time and frequency of the window function. The prerequisites of using STFT are signal stationarity within the window because burst or transient signals will significantly affect the performance. In this study, a Hann window of length $128$ ms was applied. The overlap length was $125$ ms, and the FFT was $512$ samples long.
% \textcolor{blue}{In this study, a Hann window of length $128$ ms was applied. The overlap length was $125$ ms, and the FFT was $512$ samples long.}

2)	Continuous wavelet transform \\
\indent Continuous wavelet transform (CWT) is another commonly used TFD that performs well in non-stationary signals such as EEG and ECG. It can extract better local information in both time and frequency domains. The CWT (scalogram) of signal $x(t)$ can be calculated as given in Eq. (\ref{eq:cwt}).
\begin{equation}
    \begin{aligned} \widehat{X}_{C W T} &=\frac{1}{\sqrt{\alpha}} \int_{-\infty}^{+\infty} x(t) \psi^{*}\left(\frac{t-\beta}{\alpha}\right) dt \\ &=\sqrt{\alpha} \int_{-\infty}^{+\infty} X(\omega) \psi^{*}(\alpha \omega) e^{j \omega \beta} d \omega \end{aligned}
\label{eq:cwt}
\end{equation}

\noindent where $\alpha$ is the scale parameter that is inversely related to the frequency, $\beta$ is the shifting parameter, $\psi^{*}(t)$ is the complex conjugate of the mother wavelet and $\psi(\frac{t-\beta}{\alpha})$ means the contracted and stretched mother wavelet for fitting the signal $x(t)$. In essence, the output of CWT is the convolution of the input signal with the mother wavelet. Thus, on the right-hand side equation, $X(t)$ and $\psi(\omega)$ are the Fourier transformed $x(t)$ and $\psi(t)$, respectively. Unlike STFT, which uses the same sliding window for all frequency bands, CWT can vary the window size by the scaling parameter. The mother wavelet is contracted when the scaling parameter is small, providing finer resolution. In contrast, the mother wavelet is stretched when the scaling parameter is large, providing coarser resolution. Because the scaling parameter can be regarded as the inverse frequency, the pseudo-frequency of the CWT can be approximated by Eq. (\ref{eq:pesudo-freq-cwt}):
\begin{equation}
    f=\frac{f_{mv} f_{s}}{\alpha}
\label{eq:pesudo-freq-cwt}
\end{equation}

\noindent where $f_{mv}$ is the centre frequency of the mother wavelet, $f_s$ is the sampling frequency, and $\alpha$ stands for the scale. Ergen et al. showed that Morlet is the most appropriate mother wavelet for heart sound analysis; thus, it is also selected to generate CWT \cite{ergen2012time}.
% \textcolor{blue}{Ergen et al. showed that Morlet is the most appropriate mother wavelet for heart sound analysis; thus, it is also selected to generate CWT \cite{ergen2012time}.} 

3) Chirplet transform \\
\indent Chirplet transform (CT) can be regarded as an improved wavelet transform by further modifying the mother wavelet. It rotates the 'wavelet' in the time-frequency plane, which is equivalent to applying a nonnegative, symmetric, and normalized window $\omega(\sigma)$ (usually a Gaussian window function) \cite{cui2017biosignal}. The output 'chirplet' is windowed 'wavelet' in scaling and time-shifting. In addition, another two parameters are introduced, chirping (frequency rotation operator) $\Phi_{\alpha}^{R}(t)$ and frequency shifting $\Phi_{\alpha}^{M}\left(t, t_{0}\right)$, calculated by Eq. (\ref{eq:ct}):
\begin{equation}
\begin{aligned}
    & \Phi_{\alpha}^{R}(t) = e^{-j \alpha t^{2} / 2} \\
    & \Phi_{\alpha}^{M}\left(t, t_{0}\right) = e^{j \alpha t_{0} t}
\end{aligned}
\label{eq:ct}
\end{equation}
where $j=\sqrt{-1}$, $\alpha$ is chirp rate, and $t$ is time. The rotatory angle of the analytical associate of the signal is $\theta=\tan ^{-1}(-\alpha)$ and the shift in the frequency is from $\omega$ to $\omega + \alpha t_0$. The analytical associate $z(t)$ of the signal can be obtained by Hilbert transform. The chirplet transform of a signal $x(t)$ can be expressed as Eq. (\ref{eq:ct2}):
\begin{equation}
    \hat{X}_{CT}\left(t_{0}, \omega, \alpha\right)=\int_{-\infty}^{+\infty} \bar{z}(t) W_{(\sigma)}\left(t-t_{0}\right) e^{-j \omega t} d t
\label{eq:ct2}
\end{equation}
where $\omega$ is the frequency, $\bar{z}(t)=z(t) \Phi_{\alpha}^{R}(t) \Phi_{\alpha}^{M}\left(t, t_{0}\right)$. The terminology of the rest variables is the same as above. This study applied a Gaussian window of 64 samples. The number of frequency axis points associated with the spectrum was set to 512.
% \textcolor{blue}{This study applied a Gaussian window of 64 samples. The number of frequency axis points associated with the spectrum was set to 512.}

4) Wigner-Ville distribution\\
\indent Cohen's class distributions are also known as bilinear or quadratic time-frequency distributions. Wigner–Ville distribution (WVD) is one of Cohen's class's most used distributions. As described, STFT needs to balance the time and frequency resolution; however, the bilinear time-frequency distributions, such as WVD, can offer excellent resolutions in the time and frequency domain without trade-offs. They convert the time and frequency of the signal into the complex conjugate signal to present the energy distribution, and their resolutions are only related to the signal length. The calculation of Cohen's class distribution is given by Eq. (\ref{eq:cohen_class_dist}):

\begin{equation}
\begin{split}
     & C_{x}(t, f) =  \\
     & \int_{-\infty}^{\infty} \int_{-\infty}^{\infty} A_{x}(\eta, \tau) \Phi(\eta, \tau) \exp (j 2 \pi(\eta t-\tau f)) d \eta d \tau  &
\end{split}
\label{eq:cohen_class_dist}
\end{equation}

\noindent where $A_{x}(\eta, \tau)=\int_{-\infty}^{\infty} x(t+\tau / 2) x^{*}(t-\tau / 2) e^{-j 2 \pi t_{j}} d t$ is the ambiguity function and $\Phi(\eta, \tau)$ is the kernel function (normally low-pass function) to reduce the noise. When $\Phi(\eta, \tau) = 1$ it is the WVD, expressed as in Eq. (\ref{eq:wvd}):
\begin{equation}
    \widehat{X}_{W V D}(t, f)=\int_{-\infty}^{\infty} x(t+\tau / 2) x^{*}(t-\tau / 2) e^{-j 2 \pi f \tau} d \tau
\label{eq:wvd}
\end{equation}

Although WVD provides a better resolution of the signal TFD, it has the issue of cross-term (false indication of the existence of the signal components), which is a common problem of Cohen's class distributions. Therefore, the smoothed pseudo-WVD was selected to reduce the cross-term issue in this study.

5) Choi-Williams distribution \\
\indent The formal approach for reducing and eliminating the cross-term interference of Cohen's class distributions is to select the appropriate kernel function. However, this may also corrupt the valuable information of the signal. Besides the pseudo-WVD, we also applied the Choi-Williams distribution (CWD) to allow comparison. The kernel function of the CWD is $\Phi(\eta, \tau)=e^{-\partial(\eta \tau)^{2}}$. In \cite{chen2015recognition}, they showed that when $\partial=3$, the TFD can clearly show the time, frequency, and intensity of the heart sound component.
% \textcolor{blue}{In \cite{chen2015recognition}, they showed that when $\partial=3$, the TFD can clearly show the time, frequency, and intensity of the heart sound component.}

\begin{figure}[!ht]
    \centering
    \includegraphics[width=\columnwidth]{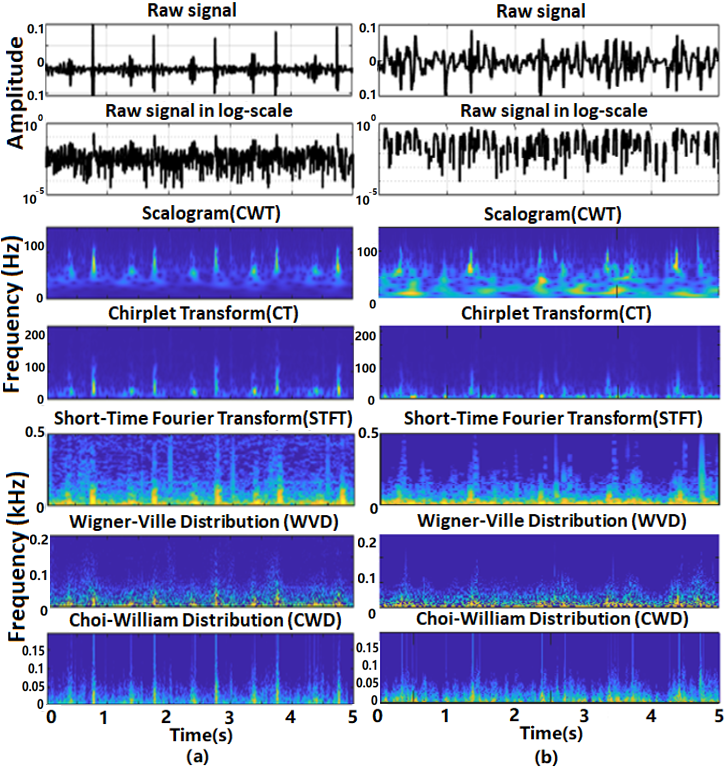}
    \caption{Visualisation of the generated TFDs. (Left) Healthy heart sound. (Right) Unhealthy heart sound.}
    \label{fig:TFDs}
\end{figure}

% \textcolor{blue}{Before the generated TFDs inputting into the CNNs, the TFD settings were also manually checked the clarity of visualisation.} 
Before the generated TFDs inputting into the CNNs, the TFD settings were also manually checked the clarity of visualisation. Fig. \ref{fig:TFDs} shows the generated TFDs from both healthy and unhealthy subjects. 

% \begin{figure*}[!ht]
%     \centering
%     \includegraphics[width=0.65\linewidth]{figures/TFDs.png}
%     \caption{Color mapping of the generated TFDs in log scale: (Left) Time-frequency distributions of the healthy heart sound. (Right) Time-frequency distributions of the unhealthy heart sound.}
%     \label{fig:TFDs}
% \end{figure*}

\subsection{CNN Models}
\label{subsec:CNNs}
CNNs have developed rapidly since AlexNet surpassed the traditional methods by a large margin in ILSVRC2012, a visual recognition competition \cite{russakovsky2015imagenet}.
CNNs have numerous variants and differ in architecture, complexity, computation load, etc. To fairly evaluate and compare the performance of CNNs in this study, four types of CNN architectures are selected, including MobileNetV3 \cite{howard2019searching}, ResNet \cite{he2016deep}, SEResNet \cite{hu2018squeeze}, and DenseNet \cite{huang2017densely}, as described below.

1) MobileNetV3 \\
\indent CNNs are composed of stacked convolutional blocks generally containing convolution, normalisation, and non-linear activation layers. The depth and width of CNNs have to be large to extract useful high-level visual features for prediction. However, the expensive computational load caused by the over-expansion of the CNN scale might be unaffordable for deploying such CNNs on either personal computers or mobile devices. Small-scale CNNs can alleviate the computational load but might decrease the network performance. Hence, efficient architecture design is crucial for CNNs to be deployed in mobile devices with satisfactory performance.

MobileNetV3, proposed in $2019$, is a lightweight CNN architecture that enables accurate and efficient computation in mobile devices for visual recognition tasks. Till now, it has been proved that MobileNetV3 can work efficiently in many computer vision tasks, e.g., image classification, segmentation, detection, etc. Its efficient and accurate computation comes from some improvements, including depth-wise separable convolution, neural architecture searching (NAS) \cite{tan2019mnasnet}, NetAdapt \cite{yang2018netadapt}, etc. In particular, depth-wise separable convolution is employed in all MobileNet series to reduce the number of trainable parameters by factorising traditional convolutions. NAS and NetAdapt are employed to optimise the network architecture at the block and layer levels. All these improvements make MobileNetV3 an efficient CNN architecture with fast interface time.

Besides, MobileNetV3 includes MobileNetV3-small and MobileNetV3-large, varying in depth and number of parameters. These two models have the fewest trainable parameters among all selected models in this study. Hence, MobileNetV3 acts as the baseline in this study for analysing the impact of model size in the heart sound classification.

2) ResNet \\
\indent Increasing the depth can enhance CNNs' capacity, but it will bring about the vanishing gradient problem when the networks are too deep. It means the gradient becomes infinitely small via the multiple multiplications during backward propagation. Consequently, when a model goes deeper, its performance reaches saturation and often drops rapidly. This problem, also known as model degradation, occurred before ResNet was proposed in 2015. 

ResNet is one of the most novel architectures in CNNs. Its essential contribution is skip-connection, which enables deep training and dramatically alleviates the degradation problem. With the operation, the function of some redundant layers turns into identity mapping when the network saturates. The gradient of a layer thus can backwards propagate directly to its previous layer without shrinking. 

Although many CNNs have been used in the heart sound classification field, there is still no evidence of whether a deeper network can improve the classification performance or decrease due to model degradation. Hence, ResNet-18/34/50 are selected in this study to analyse the network depth impact while alleviating the degradation.

3) DenseNet \\
\indent As aforementioned, ResNet builds a direct connection to allow gradient propagation between two connected layers. DenseNet makes more dense connections between all layers. In other words, the input of one layer in DenseNet contains the inputs of all its preceding layers. The dense connection improves the flow of both features and gradients, making training deep neural networks easier. Furthermore, the deeper layers in a network can extract higher-level features, while lower-level features might also benefit the classification. The dense connection in DenseNet, combining features from all layers to form multi-scale convolutional feature maps, thus might help improve the heart sound classification. Notably, our study conducts experiments on DenseNet-121/169 which is affordable in terms of computational cost.

4) SEResNet \\
\indent Squeeze-and-Excitation (SE) is a module that improves the network representations by modelling channel-wise interdependencies of convolutional features. It squeezes the convolutional features across the spatial dimension to produce a vector describing the channel-wise global distribution. The subsequent excitation assigns modulation weights for each channel to rescale the convolutional features. 

SE can easily embed to most existing CNN structures, e.g., ResNet, VGG, MobileNet, ShuffleNet, etc., and improve their performance. However, it has not been experimented with whether SE can improve CNN performance in the heart sound classification yet. We, therefore, embed it to ResNet-18/34/50 to verify if SE can help CNNs achieve higher performance in this classification task.

5) CNN Summary \\
\indent Table~\ref{tab:CNN_Summary} lists the number of trainable parameters and multiply-accumulates (MACs) of all participated CNNs. By experimenting with all these CNNs, we can analyse the relationship between model capacity and classification performance in the PCG classification. Meanwhile, MACs, representing the computational cost for a CNN to process a single signal, can help find a trade-off between classification accuracy and computational cost. Besides, experiments on four kinds of network architectures can help discover whether some of them are more appropriate for this task. 

\begin{table}[pos=!ht]
\caption{Number of parameters (in megabyte, M) and multiply-accumulates (MACs) (in gigabyte, G) of the participated CNNs.}
\label{tab:CNN_Summary}
\begin{tabular*}{\tblwidth}{@{}RRR@{}}
\toprule
Model & Params (M) & MACs (G) \\ 
\midrule
MobileNetV3-small & $1.52$ & $0.06$ \\
MobileNetV3-large & $4.20$ & $0.22$ \\
ResNet18 & $11.18$ & $1.82$ \\
ResNet34 & $21.29$ & $3.67$ \\
ResNet50 & $23.51$ & $4.11$ \\
SEResNet18 & $11.27$ & $1.82$ \\
SEResNet34 & $21.45$ & $3.67$ \\
SEResNet50 & $26.04$ & $4.11$ \\
DenseNet121 & $6.87$ & $2.83$ \\
DenseNet169 & $12.33$ & $3.36$ \\
\bottomrule
\end{tabular*}
\end{table}

\subsection{Datasets \& Pre-processing}
\label{subsec:datasets}
The database used in this study was the PhysioNet database which consists of $3153$ recordings, including $2488$ normal and $665$ abnormal cases. They were recorded by different teams using different electronic stethoscopes in both clinical and non-clinical settings. Because of the uncontrolled environment, the recording durations ranged from $5$s to $120$s. The subjects included children, adults, and the elderly. The abnormal cases involved various heart conditions, but the majority were coronary and valvular diseases (read more in \cite{liu2016open}).

Referring to \cite{bao2022effect}, the recordings were further segmented into $5$s durations without pre-processing overlapping, generating $13015$ segments ($9857$ normal and $3158$ abnormal). Since the testing datasets were not included in the published Physionet challenge database, we randomly divided the generated segments into training, validation, and testing sets by $8:1:1$. In all three sets, the distribution of the normal and abnormal segments retained approximately $3:1$.

\subsection{Training Settings}
\label{subsec:training_settings}
This study kept all the hyper-parameters consistent in all experiments for a fair comparison. The computer used one Nvidia Tesla V100 GPU card. The CNN training codes are implemented by Pytorch 1.9 \cite{paszke2019pytorch}. All CNN backbones were implemented by the Timm library \cite{rw2019timm}. The training procedure was accelerated by automatic mixed precision (AMP) in Pytorch to save the computing memory. The results were obtained by averaging the measurements in $10$ repeated experiments with different random seeds to make the comparison more convincing and alleviate randomness concerns.

The input signal images for CNNs were resized to a standard resolution of $224\times224$ in three channels (RGB), and the pixel values were normalised with zero mean and unit standard deviation. For raw signals and their log-scale form, their 2-D images were generated by projecting the sequential signal amplitude in y-axis, connecting neighboring points (equivalent to the “Plot” function), getting frame to obtain the waveform, and resizing the figure to standard resolution. Both single TFD and their combination were tested as CNN inputs. Specifically, we loaded three TFDs in greyscale and stacked the three greyscale inputs in channel dimension. The dimension of the combined TFDs is consistent with that of one single TFD, so no modification to the network architectures to fit the input size was needed.
% \textcolor{blue}{For raw signals and their log-scale form, their 2-D images were generated by projecting the sequential signal amplitude in y-axis, connecting neighboring points (equivalent to the “Plot” function), getting frame to obtain the waveform, and resizing the figure to standard resolution.} 

The employed optimiser is Adamw with $10^{-2}$ weight decay.
According to the cosine annealing decay schedule, the learning rate decays from an initial $10^{-2}$ to the minimum $10^{-6}$ in $50$ epochs. The batch size is set to $128$ to ensure batch variety and avoid memory overflow.

The performance of TFDs was compared with the baseline using original or log-scaled raw signal as CNN inputs.

\subsection{Performance Metrics}
\label{subsec:metrics}
Accuracy is the key metric for evaluating the performance of a classification algorithm. However, the data structure in this study was not balanced (normal: abnormal is approximately $3: 1$), so the true positive rate (sensitivity, Se), true negative rate (specificity, Sp), and overall score (MAcc) were also calculated as in Eq. (\ref{eq:metrics}) to obtain balanced results.
\begin{equation}
% \begin{aligned}
% Acc= & \frac{TP+TN}{TP+FP+TN+FN} \\
% Se= & \frac{TP}{TP+FN} \\
% Sp= & \frac{TN}{TN+FP} \\
% MAcc= & \frac{Se+Sp}{2} \\
% \end{aligned}
\begin{split}
    & Acc= \frac{TP+TN}{TP+FP+TN+FN} & Se=   \frac{TP}{TP+FN} \\
    & Sp=  \frac{TN}{TN+FP}          & MAcc= \frac{Se+Sp}{2} \\
\end{split}
\label{eq:metrics}
\end{equation}

% where TP (True positive) is the correctly classified healthy condition cases and TN (True negative) indicates the correctly classified unhealthy cases. Similarly, FP (False positive) represents the false detection on the normal sets and FN (False Negative) means incorrectly identified abnormal cases. The overall score (MAcc) is the average of the Se and Sp.

where TP (True Positive) and TN (True Negative) denote the correctly classified healthy and unhealthy condition cases, respectively. Similarly, FP (False Positive) and FN (False Negative) represents the false detection on the normal and abnormal sets, respectively. The overall score (MAcc) is the average of the Se and Sp.

\subsection{Statistical Analysis}
\label{subsec:statis_ana}
All CNNs were trained and tested ten times with different TFDs. In particular, the CNN training processes were set to be deterministic but with ten random seeds. The non-parametric test (Mann-Whitney U test) was conducted for statistical significance based on the experimental results. The p-value of the test can reveal whether a TFD or CNN outperforms the other objectively. If the p-value is more significant than $0.05$, there is no significant difference between two TFDs or CNNs. 
\section{Results}
\label{sec:results}

\subsection{TFD effects on the CNN performance}
\label{subsec:effect_tfds}
\begin{figure}[!ht]
    \centering
    \begin{subfigure}{\columnwidth} \centering
        \includegraphics[width=0.75\columnwidth]{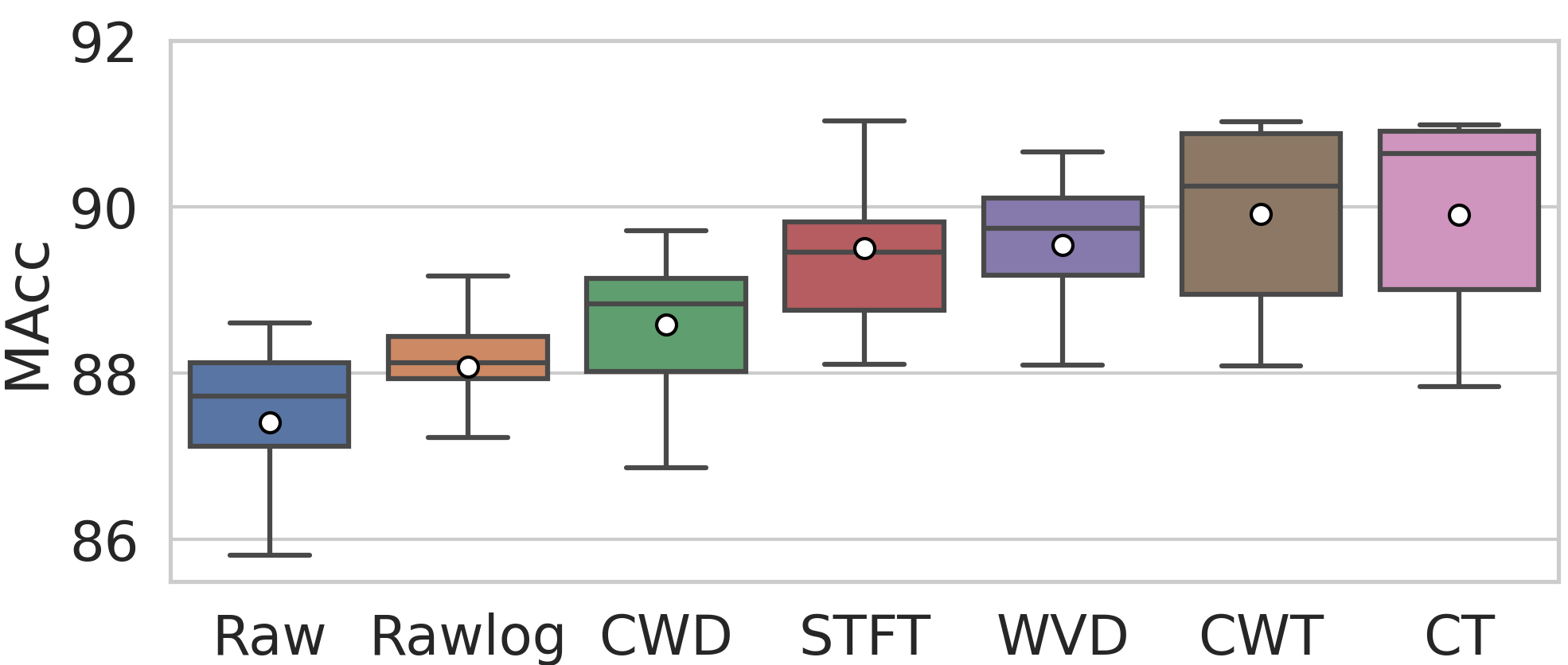}
        \caption{}
    \label{subfig:signal_effect_overall}
    \end{subfigure}
    \begin{subfigure}{\columnwidth} \centering
        \includegraphics[width=0.75\columnwidth]{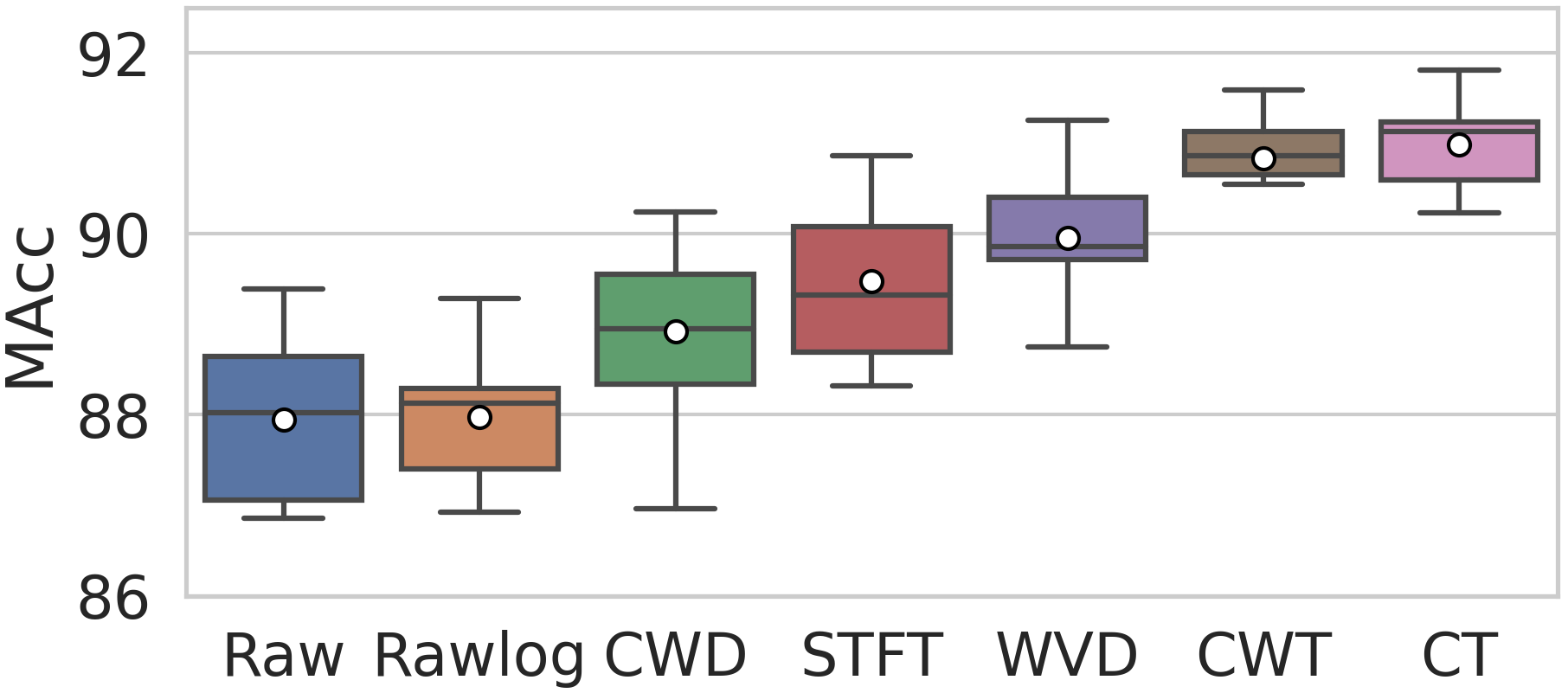}
        \caption{}
        \label{subfig:signal_effect_resnet18}
    \end{subfigure}
    \begin{subfigure}{\columnwidth} \centering
        \includegraphics[width=0.75\columnwidth]{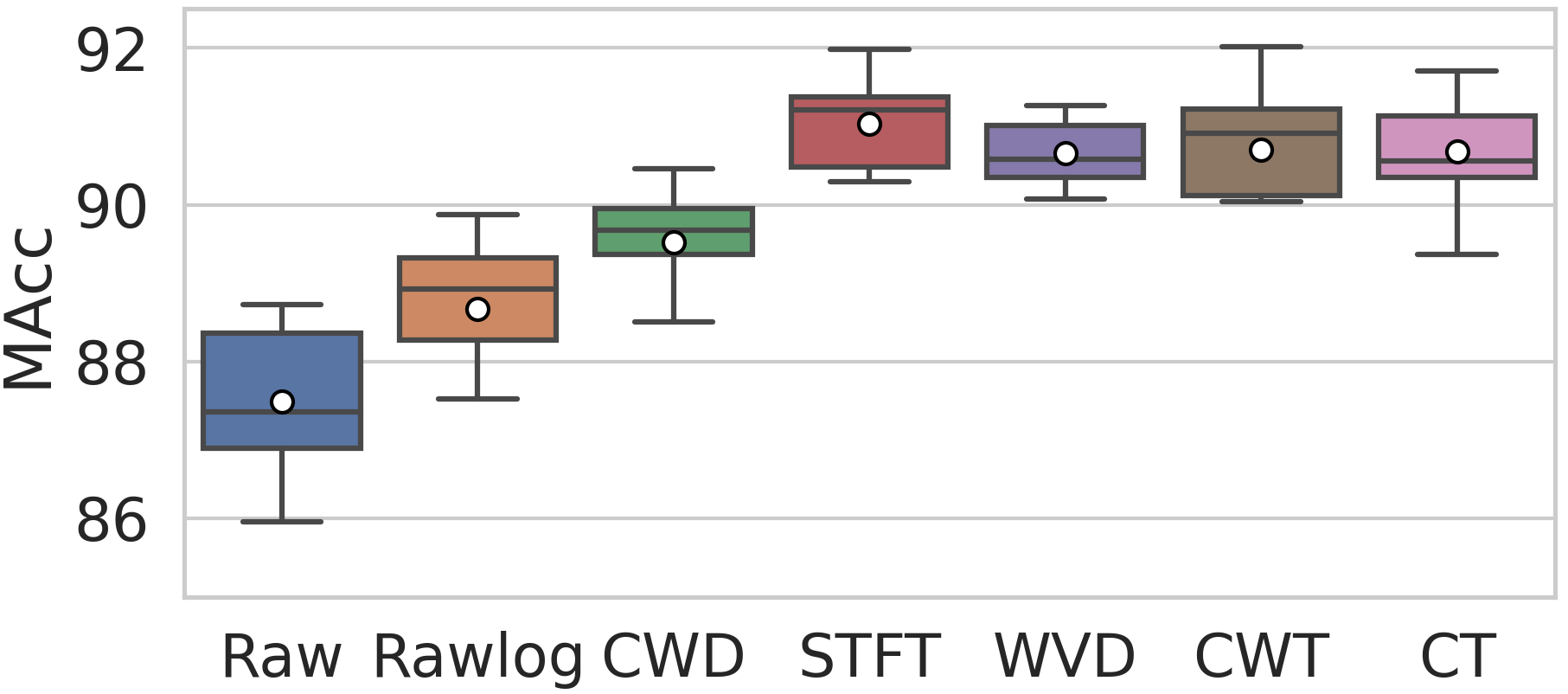}
        \caption{}
        \label{subfig:signal_effect_dense121}
    \end{subfigure}
    \caption{(a) Overall performances of TFDs on all $10$ CNNs, (b) Performances of TFDs on ResNet18, (c) Performances of TFDs on DenseNet121. The horizontal lines and the in-box white dots represent the median and average values, respectively.}
    \label{fig:signal_effect}
\end{figure}

The overall performances (in MAcc, an average of $10$ times) of the TFDs for $10$ CNNs are illustrated in Fig. \ref{subfig:signal_effect_overall}. It revealed that transforming from the 1-D time domain to the 2-D time-frequency domain by TFDs can stably improve the classification performance. In particular, compared with the baseline (raw signal, $87.4\%$), the transform can improve the overall performance by up to $2.5\%$. Among the five TFDs, CWT and CT achieved around $89.9\%$ average MAcc, surpassing the others by approximately $0.5-1.3\%$ STFT and WVD achieved comparable MAcc of around $89.5\%$, slightly worse than CWT and CT. CWD performed the worst with $88.6\%$. Besides, it is worth noting that using a log-scaled raw signal can gain higher performance than the baseline raw signal, improving from $87.4\%$ to $88.1\%$ approximately.

Overall, CWT and CT were the two superior TFDs in our experiments, and there was no statistical difference between them ($p=0.815$). STFT and WVD followed with trivial performance differences ($p=0.413$). The p-values between CWT/CT and STFT/WVD were in the range of $0.012$ to $0.035$ in pair, indicating CWT/CT outperformed STFT/WVD statistically. CWD achieved slightly worse classification performance among the five TFDs, and its p-value concerning any other TFD was more petite than $0.001$, indicating significant inferiority. Likewise, the baseline raw signal performed with the lowest MAcc, and its p-values concerning the others are all infinitely close to $0$, indicating that using TFDs instead of the raw signal can improve the classification performance stably. Besides, the log-scaled raw signal achieved an intermediate performance between the baseline and the TFDs. Moreover, this can be proved objectively that its corresponding p-values approached $0$ infinitely. 

As aforementioned, the overall performance differences were found among the TFDs. It is worth noting that the differences are CNN model dependent. For example, Fig. \ref{subfig:signal_effect_resnet18} and Fig. \ref{subfig:signal_effect_dense121} demonstrated the performances of ResNet18 and DenseNet121, respectively. It can be observed that the difference ($>1\%$) between STFT/WVD (MAcc: $89.47\%$ / $89.96\%$) and CWT/CT (MAcc: $90.84\%$ / $90.99\%$) existed for ResNet18. However, for DenseNet121, their performances were all around $90.8\%$, with differences smaller than $0.4\%$.

\subsection{Comparison of the CNN performances}
\label{subsec:CNN_effect}
\begin{figure}[!ht]
    \centering
    \begin{subfigure}{\columnwidth} \centering
        \includegraphics[width=\columnwidth]{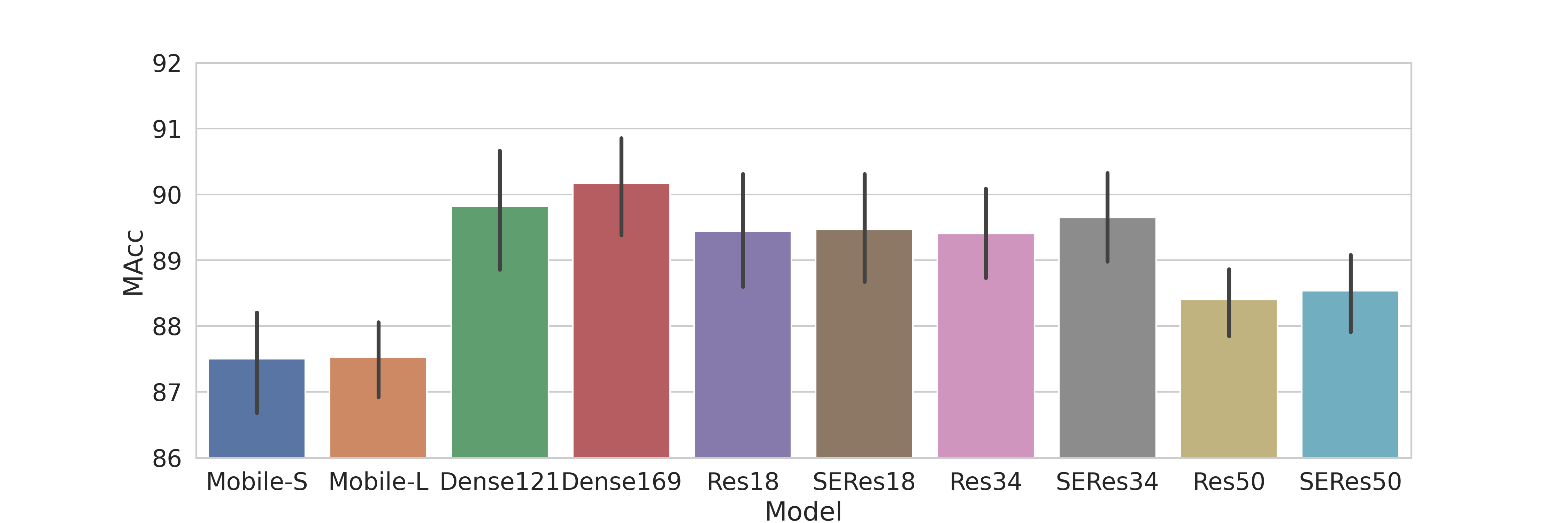}
        \caption{}
    \label{subfig:model_effect_overall}
    \end{subfigure}
    \begin{subfigure}{\columnwidth} \centering
        \includegraphics[width=.9\columnwidth]{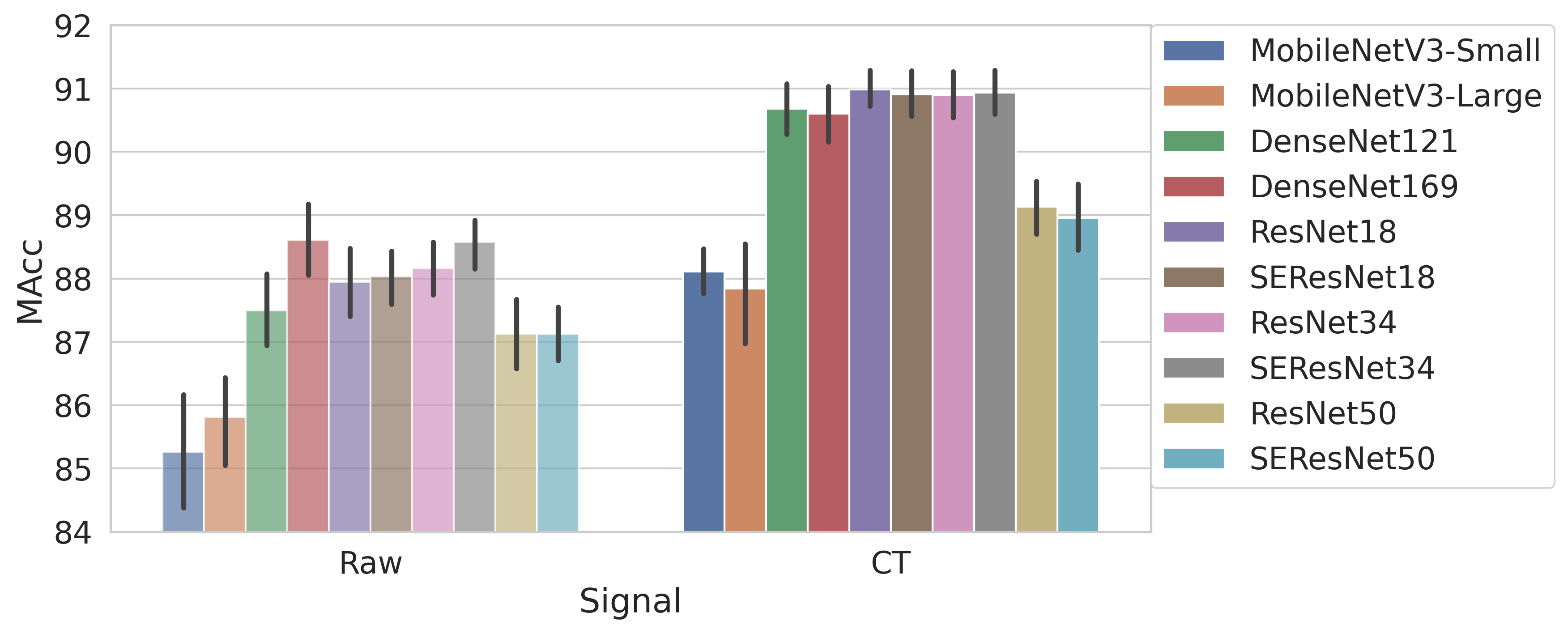}
        \caption{}
        \label{subfig:model_effect_raw_ct}
    \end{subfigure}
    \begin{subfigure}{\columnwidth} \centering
        \includegraphics[width=0.8\columnwidth]{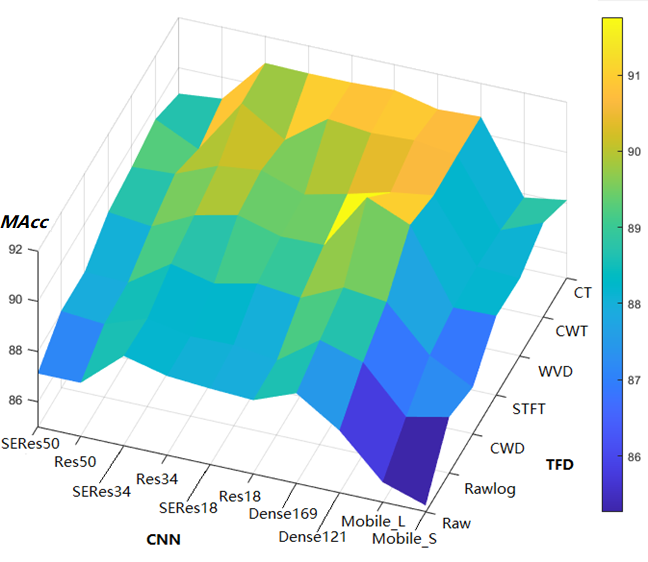}
        \caption{}
        \label{subfig:performance_10CNNs}
    \end{subfigure}
    \caption{(a) Overall performances of $10$ CNNs with all seven input types, (b) Performances of all the $10$ CNNs on raw signal and CT. (c) Performances of the $10$ CNNs concerning the seven TFDs.}
    \label{fig:model_effect}
\end{figure}

The overall performances (in MAcc, an average of $10$ times) of $10$ CNNs with all seven input types were illustrated in Fig. \ref{subfig:model_effect_overall}. The models were sorted by the number of parameters from left to right. It can be observed that the lightweight MobileNets did not perform well, achieving approximately $87.50\%$ MAcc. On the other hand, the performances of ResNet-18/34, SEResNet-18/34, and DenseNet121 were quite close, around $89.6\%$ MAcc with differences below $0.4\%$. DenseNet169 outperformed all the others slightly and achieved $90.17\%$ MAcc. Besides, ResNet50 and SEResNet50 using bottleneck block architecture achieved an unsatisfactory performance of around $88.5\%$.

To compare the performance of different CNNs, a statistical analysis was conducted. Particularly, the p-value between MobileNet-Small and MobileNet-Large was $0.791$, which means they have no statistical difference. The p-values between ResNet-18/34 and SEResNet-18/34 were above $0.05$ (in the range from $0.313$ to $0.987$), implying no statistical difference. The p-values of DenseNet121 concerning the four CNNs mentioned above ranged from $0.087$ to $0.193$, except for ResNet34 ($0.031$). This showed that DenseNet121 had a statically close performance with ResNet18 and SEResNet18/34 but was slightly better than ResNet34. DenseNet169 achieved relatively higher performance than the others, with p-values smaller than $0.017$. Besides the CNNs mentioned above, ResNet50 and SEResNet50 shared close performance with a p-value of $0.357$, higher than MobileNets but lower than the others. 

Besides the overall performance of $10$ CNNs, raw signal and CT performances were visualised on these CNNs as presented in Fig. \ref{subfig:model_effect_raw_ct}. The visualisation revealed the difference between raw signal and CT but presented different relationships between CNNs. For example,
when using raw signal, DenseNet169 achieved the highest performance, $88.60\%$. But when using CT, it performed slightly worse than ResNet-18/34 or SEResNet-18/34 (DenseNet169: $90.60\%$, others: approximately $90.90\%$). Fig. \ref{subfig:performance_10CNNs} represents the overall performance using a surface plot.

\subsection{Effect of combined TFDs as CNN Input}
\label{subsec:combination}
As described in Section \ref{subsec:effect_tfds}, there was no apparent difference in the CNN performance using the TFDs except that CWD performed slightly worse. Since the TFDs transform the raw signals into different time-frequency domains, it is meaningful to analyse whether combining TFDs can improve CNN performances. Therefore, experiments were conducted on various channel-wise combinations of different TFDs as the inputs for CNNs. Notably, three CNNs representing three different network architectures with higher depth and better performance in Section \ref{subsec:CNN_effect} were selected, including ResNet34, SEResNet34 and DenseNet169, to analyze the effect of the TFD combinations. MobileNets were not included because of the slightly worse performance and shallower architectures had no advantage in fitting more information. Although SE/ResNet50 were deeper in architectures, they were not selected due to poorer performance compared with SE/ResNet34.
% \textcolor{blue}{Notably, three CNNs representing three different network architectures with higher depth and better performance in Section \ref{subsec:CNN_effect} were selected, including ResNet34, SEResNet34 and DenseNet169, to analyze the effect of the TFD combinations. MobileNets were not included because of the slightly worse performance and shallower architectures had no advantage in fitting more information. Although SE/ResNet50 were deeper in architectures, they were not selected due to poorer performance compared with SE/ResNet34.}
As aforementioned in Section \ref{subsec:training_settings}, three out of the four TFDs (except CWD) are stacked in channel dimension to form the combined TFDs. Table~\ref{tab:combination} lists the performances ($10$ times average, sensitivity, specificity, and MAcc, respectively) of three CNNs using different single or combined TFDs. The results showed that combining input did not improve classification performance.

\begin{table}[pos=!ht]
\caption{Comparison of classification performance ($\%$) between combined TFDs and single TFD (an average of $10$ times).}
\label{tab:combination}
\begin{tabular*}{\tblwidth}{@{}CCCCCCCC@{}}
\toprule
\multirow{2}{*}{Model} & \multicolumn{4}{c}{Combined vs. Single} & \multirow{2}{*}{Se} & \multirow{2}{*}{Sp} & \multirow{2}{*}{MAcc} \\
 & CT & CWT & STFT & WVD & & &  \\
 \midrule
\multirow{8}{*}{ResNet34} & \checkmark & & & & $85.83$ & $95.97$ & $90.90$ \\
 & & \checkmark & & & $83.85$ & $95.97$ & $89.80$ \\
 & & & \checkmark & & $84.17$ & $95.65$ & $89.91$ \\
 & & & & \checkmark & $84.81$ & $95.49$ & $90.15$ \\
 & \checkmark & \checkmark & \checkmark & & $85.51$ & $95.24$ & $90.38$ \\
 & \checkmark & \checkmark & & \checkmark & $86.11$ & $95.37$ & $90.74$ \\
 & \checkmark & & \checkmark & \checkmark & $83.44$ & $94.94$ & $89.19$ \\
 & & \checkmark & \checkmark & \checkmark & $85.80$ & $95.41$ &	$90.61$ \\
 \midrule
\multirow{8}{*}{SEResNet34} & \checkmark & & & & $85.61$ & $96.27$ & $90.94$ \\
 & & \checkmark & & & $85.86$ & $95.91$ & $90.89$ \\
 & & & \checkmark & & $83.50$ & $95.56$ & $89.53$ \\
 & & & & \checkmark & $84.33$ & $95.56$ & $89.94$ \\
 & \checkmark & \checkmark & \checkmark & & $86.50$ & $95.35$ & $90.93$ \\
 & \checkmark & \checkmark & & \checkmark & $86.31$ & $95.91$ & $91.11$ \\
 & \checkmark & & \checkmark & \checkmark & $83.85$ & $95.32$ & $89.59$ \\
 & & \checkmark & \checkmark & \checkmark & $84.52$ & $95.50$ &	$90.01$ \\
\midrule
\multirow{8}{*}{DenseNet169} & \checkmark & & & & $85.76$ & $95.44$ & $90.60$ \\
 & & \checkmark & & & $86.11$ & $95.73$ & $90.92$ \\
 & & & \checkmark & & $\mathbf{87.29}$ & $96.22$ & $\mathbf{91.76}$ \\
 & & & & \checkmark & $85.25$ & $95.56$ & $90.41$ \\
 & \checkmark & \checkmark & \checkmark & & $85.92$ & $\mathbf{96.37}$ & $91.14$ \\
 & \checkmark & \checkmark & & \checkmark & $85.57$ & $96.29$ & $90.93$ \\
 & \checkmark & & \checkmark & \checkmark & $85.89$ & $95.79$ & $90.84$ \\
 & & \checkmark & \checkmark & \checkmark & $86.66$ & $96.31$ &	$91.48$ \\
\bottomrule
\end{tabular*}
\end{table}

\section{Discussion}
\label{sec:discussion}
This study compared several commonly used TFDs to classify the heart sound signals on ten mainstream CNNs and explored using combined TFDs as inputs to improve the classification performance. In previous heart sound analysis studies, TFDs were only discussed for providing better human visualisation or more comprehensive information. They were used individually as CNN inputs in the classification, while no study has objectively compared and analysed their performance under different CNNs. Computer-aided analysis for heart sound is the future trend that shows the potential to overcome the physician's clinical experience and listening limitation, providing a more accurate and robust diagnosis. Thus, it is reasonably necessary to know if and how the selection of the TFDs will affect the CNN performance. This study filled this gap and researched the input and model selection. The results revealed that using log-scale for the raw signal as input in the time domain can improve classification performance with approximately $1-2\%$ in MAcc on MobileNetV3 but a trivial increase ($<1\%$) on deeper networks. On almost all the models, converting the heart sound signal from the time domain into the time-frequency domain by TFDs can improve classification performance by approximately $2.3\%$ in MAcc.

Furthermore, the selection of TFDs was not decisive on the heart sound classification, but it will still affect the performance. As illustrated in Section \ref{subsec:effect_tfds}, where we compared all TFDs, CT and CWT are the two superior TFDs that outperformed STFT and WVD by a small margin and outperformed CWD. We assume that the slightly worse performance ($<1\%$ than the rest TFDs) of CWD is due to the parameter setting $\partial$ of its kernel function. Although \cite{chen2015recognition} claimed that $\partial=3$ could provide better resolution, it may filter out useful information.
% Though in \cite{chen2015recognition}, it was claimed that $\partial=3$ can provide better resolution. However, it may filter useful information. 
Moreover, due to the different heart sound recording settings, a suitable $\partial$ may be further discussed. Even though TFDs were obtained by different mathematical calculations with different visual representations, as shown in Fig. \ref{fig:TFDs}, the hidden information provided was similar. Their morphological differences decrease significantly as CNN input with lower resolution ($224\times224$), resulting in similar classification performances. Nevertheless, results suggest that CWT is more suitable than the rest TFDs as input for heart sound classification.

Secondly, the performances of different CNNs concerning TFDs were compared. It has been observed that the increase of model capacity was necessary for classifying the 2D inputs since the lightweight MobileNetV3-small/large performed worse than the other models with more parameters. However, only increasing the depth or the parameter number of CNNs without changing architecture does not guarantee a better result. This can be observed from ResNet and SEResNet family as in Fig. \ref{fig:model_effect}, that their classification performance remains the same when increasing depth from $18$ to $34$ (within $0.15\%$ difference in MAcc, $p\leq 0.313$). Meanwhile, their performance significantly drops when increasing the depth from $34$ to $50$ for the two CNNs. This drop may come from the convolutional block. It changes from basic to bottleneck, where the convolutional features lose effectivity in deeper layers. SE was an approach to improve network performance, while there was no observed improvement in this study. This may be due to no extra valuable features were extracted by squeezing the convolutional features across the spatial dimension. 

Furthermore, DenseNet-121/169 with fewer parameters achieve comparable performance to ResNet or SEResNet. It indicates that the dense layer-wise connection in DenseNet combining multi-scale convolutional features might benefit the heart sound classification. Based on the findings above, we conclude that increasing CNN capacity appropriately and extracting multi-scale convolutional features are two significant factors for improving CNNs performance in classifying 2D heart sound signals. Although MobileNetV3 achieved lower performance than the other more extensive networks ($<3\%$ lower in MAcc), it costs much less computational resources as listed in Table~\ref{tab:CNN_Summary}. This indicates that some tiny networks like MobileNetV3 can perform heart sound classification in portable devices. In contrast, the other more extensive networks are more suitable for deploying in medical institutes or high-performance computing devices. 

Thirdly, we explored the TFD combinations on three CNN models, including ResNet34, SEResNet34, and DenseNet169. The TFDs selected included CT, CWT, STFT, and WVD, while CWD was excluded due to its worse performance. The combination was implemented by placing three out of the four selected TFDs in the three channels of the input image for CNNs. The results in Table~\ref{tab:combination} indicate that the combination methods did not yield better performance than using single TFD for all selected CNNs. The lack of improvement can be explained by the fact that the TFD representations as CNN inputs were similar and sometimes overlapped; thus, the combined TFDs cannot provide extra helpful information than a single TFD.
% \textcolor{blue}{The results in Table~\ref{tab:combination} indicate that the combination methods did not yield better performance than using single TFD for all selected CNNs. The lack of improvement can be explained by the fact that the TFD representations as CNN inputs were similar and sometimes overlapped; thus, the combined TFDs cannot provide extra helpful information than a single TFD.}
Although the combination of TFDs did not improve the heart sound classification performance visibly, the performance of the CNNs and TFDs in this study is still considerable. Compared with rank $1$ ($0.8602$ in MAcc) in the $2016$ PhysioNet/ Computing in Cardiology (CinC) Challenge, the performance in this study (approximately $0.9$ in MAcc) is higher. This showed that using the appropriate TFD and deep learning model is the optimum approach for heart sound classification.

\section{Conclusion}
\label{sec:conclusion}
This study revealed that when using CNN models to identify heart sound, it is beneficial to convert the raw signal from the time domain into the time-frequency domain, providing more comprehensive information to improve classification performance. Using TFDs can alleviate the need for training deeper CNNs to achieve identical performance. The effect of selecting time-frequency representations is not apparent for CNNs though they are different in human visualisation; we recommend CWT due to its stable performance in this study, generality, and understandability. For the CNN models, small-scale CNNs cannot perform as well as large-scale models (approximately $2-3\%$ less in MAcc), but they showed potential for embedded or portable devices. Among the mainstreamed large-scale CNN models, only increasing the depth or the parameter number is unnecessary for performance improvement and may even cause effectivity loss. The exploration of combining TFDs in the channel dimension as CNN input did not show improvement. 
\printcredits
\section*{Acknowledgement}
This study was supported in part by the King’s-China Scholarship Council PhD Scholarship programme, and has been funded in part from KFAS, Kuwait Foundation for Advancement of Sciences, project no. CN20-13EE-01. This work used the Cirrus UK National Tier-2 HPC Service at EPCC (\url{http://www.cirrus.ac.uk}) funded by the University of Edinburgh and EPSRC (EP/P020267/1).
\section*{Declaration}
The authors declare no conflicts of interest.

%% Loading bibliography style file
\bibliographystyle{model1-num-names}

% Loading bibliography database
\bibliography{references.bib}

\begin{thebibliography}{62}
\expandafter\ifx\csname natexlab\endcsname\relax\def\natexlab#1{#1}\fi
\providecommand{\url}[1]{\texttt{#1}}
\providecommand{\href}[2]{#2}
\providecommand{\path}[1]{#1}
\providecommand{\DOIprefix}{doi:}
\providecommand{\ArXivprefix}{arXiv:}
\providecommand{\URLprefix}{URL: }
\providecommand{\Pubmedprefix}{pmid:}
\providecommand{\doi}[1]{\href{http://dx.doi.org/#1}{\path{#1}}}
\providecommand{\Pubmed}[1]{\href{pmid:#1}{\path{#1}}}
\providecommand{\bibinfo}[2]{#2}
\ifx\xfnm\relax \def\xfnm[#1]{\unskip,\space#1}\fi
%Type = Book
\bibitem[{Mendis et~al.(2011)Mendis, Puska, Norrving, Organization
  et~al.}]{mendis2011global}
\bibinfo{author}{S.~Mendis}, \bibinfo{author}{P.~Puska}, \bibinfo{author}{B.~e.
  Norrving}, \bibinfo{author}{W.~H. Organization}, et~al.,
  \bibinfo{title}{Global atlas on cardiovascular disease prevention and
  control}, \bibinfo{publisher}{World Health Organization},
  \bibinfo{year}{2011}.
%Type = Misc
\bibitem[{{World Health Organization}(2021)}]{who2021cvds}
\bibinfo{author}{{World Health Organization}}, \bibinfo{title}{Cardiovascular
  diseases ({CVDs})},
  \bibinfo{howpublished}{\url{https://www.who.int/news-room/fact-sheets/detail/cardiovascular-diseases-(cvds)}},
  \bibinfo{year}{2021}.
%Type = Article
\bibitem[{Kumar and Thompson(2013)}]{kumar2013evaluation}
\bibinfo{author}{K.~Kumar}, \bibinfo{author}{W.~R. Thompson},
\newblock \bibinfo{title}{Evaluation of cardiac auscultation skills in
  pediatric residents},
\newblock \bibinfo{journal}{Clinical pediatrics} \bibinfo{volume}{52}
  (\bibinfo{year}{2013}) \bibinfo{pages}{66--73}.
%Type = Article
\bibitem[{Lam et~al.(2005)Lam, Lee, Boey, Ng, Hey, Ho, and
  Cheong}]{lam2005factors}
\bibinfo{author}{M.~Lam}, \bibinfo{author}{T.~Lee}, \bibinfo{author}{P.~Boey},
  \bibinfo{author}{W.~Ng}, \bibinfo{author}{H.~Hey}, \bibinfo{author}{K.~Ho},
  \bibinfo{author}{P.~Cheong},
\newblock \bibinfo{title}{Factors influencing cardiac auscultation proficiency
  in physician trainees},
\newblock \bibinfo{journal}{Singapore medical journal} \bibinfo{volume}{46}
  (\bibinfo{year}{2005}) \bibinfo{pages}{11}.
%Type = Article
\bibitem[{Gerbarg et~al.(1963)Gerbarg, Taranta, Spagnuolo, and
  Hofler}]{gerbarg1963computer}
\bibinfo{author}{D.~S. Gerbarg}, \bibinfo{author}{A.~Taranta},
  \bibinfo{author}{M.~Spagnuolo}, \bibinfo{author}{J.~J. Hofler},
\newblock \bibinfo{title}{Computer analysis of phonocardiograms},
\newblock \bibinfo{journal}{Progress in Cardiovascular Diseases}
  \bibinfo{volume}{5} (\bibinfo{year}{1963}) \bibinfo{pages}{393--405}.
%Type = Inproceedings
\bibitem[{Bobillo(2016)}]{bobillo2016tensor}
\bibinfo{author}{I.~J.~D. Bobillo},
\newblock \bibinfo{title}{A tensor approach to heart sound classification},
\newblock in: \bibinfo{booktitle}{2016 Computing in Cardiology Conference
  (CinC)}, \bibinfo{organization}{IEEE}, \bibinfo{year}{2016}, pp.
  \bibinfo{pages}{629--632}.
%Type = Article
\bibitem[{Amiri and Armano(2013)}]{amiri2013intelligent}
\bibinfo{author}{A.~M. Amiri}, \bibinfo{author}{G.~Armano},
\newblock \bibinfo{title}{An intelligent diagnostic system for congenital heart
  defects},
\newblock \bibinfo{journal}{Editorial Preface} \bibinfo{volume}{4}
  (\bibinfo{year}{2013}) \bibinfo{pages}{93--98}.
%Type = Article
\bibitem[{Safara et~al.(2013)Safara, Doraisamy, Azman, Jantan, and
  Ramaiah}]{safara2013multi}
\bibinfo{author}{F.~Safara}, \bibinfo{author}{S.~Doraisamy},
  \bibinfo{author}{A.~Azman}, \bibinfo{author}{A.~Jantan},
  \bibinfo{author}{A.~R.~A. Ramaiah},
\newblock \bibinfo{title}{Multi-level basis selection of wavelet packet
  decomposition tree for heart sound classification},
\newblock \bibinfo{journal}{Computers in biology and medicine}
  \bibinfo{volume}{43} (\bibinfo{year}{2013}) \bibinfo{pages}{1407--1414}.
%Type = Inproceedings
\bibitem[{Jaramillo-Garzon et~al.(2008)Jaramillo-Garzon, Quiceno-Manrique,
  Godino-Llorente, and Castellanos-Dominguez}]{jaramillo2008feature}
\bibinfo{author}{J.~Jaramillo-Garzon}, \bibinfo{author}{A.~Quiceno-Manrique},
  \bibinfo{author}{I.~Godino-Llorente}, \bibinfo{author}{C.~G.
  Castellanos-Dominguez},
\newblock \bibinfo{title}{Feature extraction for murmur detection based on
  support vector regression of time-frequency representations},
\newblock in: \bibinfo{booktitle}{2008 30th Annual International Conference of
  the IEEE Engineering in Medicine and Biology Society},
  \bibinfo{organization}{IEEE}, \bibinfo{year}{2008}, pp.
  \bibinfo{pages}{1623--1626}.
%Type = Inproceedings
\bibitem[{Singh-Miller and Singh-Miller(2016)}]{singh2016using}
\bibinfo{author}{N.~E. Singh-Miller}, \bibinfo{author}{N.~Singh-Miller},
\newblock \bibinfo{title}{Using spectral acoustic features to identify abnormal
  heart sounds},
\newblock in: \bibinfo{booktitle}{2016 Computing in Cardiology Conference
  (CinC)}, \bibinfo{organization}{IEEE}, \bibinfo{year}{2016}, pp.
  \bibinfo{pages}{557--560}.
%Type = Article
\bibitem[{Liu et~al.(2012)Liu, Wang, Liu, and Zhang}]{liu2012autonomous}
\bibinfo{author}{J.~Liu}, \bibinfo{author}{H.~Wang}, \bibinfo{author}{W.~Liu},
  \bibinfo{author}{J.~Zhang},
\newblock \bibinfo{title}{Autonomous detection and classification of congenital
  heart disease using an auscultation vest},
\newblock \bibinfo{journal}{Journal of Computational Information Systems}
  \bibinfo{volume}{8} (\bibinfo{year}{2012}) \bibinfo{pages}{485--492}.
%Type = Article
\bibitem[{Patidar et~al.(2015)Patidar, Pachori, and
  Garg}]{patidar2015automatic}
\bibinfo{author}{S.~Patidar}, \bibinfo{author}{R.~B. Pachori},
  \bibinfo{author}{N.~Garg},
\newblock \bibinfo{title}{Automatic diagnosis of septal defects based on
  tunable-q wavelet transform of cardiac sound signals},
\newblock \bibinfo{journal}{Expert Systems with Applications}
  \bibinfo{volume}{42} (\bibinfo{year}{2015}) \bibinfo{pages}{3315--3326}.
%Type = Article
\bibitem[{Wang et~al.(2007)Wang, Lim, Chauhan, Foo, and
  Anantharaman}]{wang2007phonocardiographic}
\bibinfo{author}{P.~Wang}, \bibinfo{author}{C.~S. Lim},
  \bibinfo{author}{S.~Chauhan}, \bibinfo{author}{J.~Y.~A. Foo},
  \bibinfo{author}{V.~Anantharaman},
\newblock \bibinfo{title}{Phonocardiographic signal analysis method using a
  modified hidden markov model},
\newblock \bibinfo{journal}{Annals of Biomedical Engineering}
  \bibinfo{volume}{35} (\bibinfo{year}{2007}) \bibinfo{pages}{367--374}.
%Type = Article
\bibitem[{Dokur and {\"O}lmez(2008)}]{dokur2008heart}
\bibinfo{author}{Z.~Dokur}, \bibinfo{author}{T.~{\"O}lmez},
\newblock \bibinfo{title}{Heart sound classification using wavelet transform
  and incremental self-organizing map},
\newblock \bibinfo{journal}{Digital Signal Processing} \bibinfo{volume}{18}
  (\bibinfo{year}{2008}) \bibinfo{pages}{951--959}.
%Type = Article
\bibitem[{Phanphaisarn et~al.(2011)Phanphaisarn, Roeksabutr, Wardkein,
  Koseeyaporn, and Yupapin}]{phanphaisarn2011heart}
\bibinfo{author}{W.~Phanphaisarn}, \bibinfo{author}{A.~Roeksabutr},
  \bibinfo{author}{P.~Wardkein}, \bibinfo{author}{J.~Koseeyaporn},
  \bibinfo{author}{P.~Yupapin},
\newblock \bibinfo{title}{Heart detection and diagnosis based on {ECG} and
  {EPCG} relationships},
\newblock \bibinfo{journal}{Medical devices (Auckland, NZ)} \bibinfo{volume}{4}
  (\bibinfo{year}{2011}) \bibinfo{pages}{133}.
%Type = Article
\bibitem[{Ari and Saha(2009)}]{ari2009search}
\bibinfo{author}{S.~Ari}, \bibinfo{author}{G.~Saha},
\newblock \bibinfo{title}{In search of an optimization technique for artificial
  neural network to classify abnormal heart sounds},
\newblock \bibinfo{journal}{Applied Soft Computing} \bibinfo{volume}{9}
  (\bibinfo{year}{2009}) \bibinfo{pages}{330--340}.
%Type = Inproceedings
\bibitem[{Zeyer et~al.(2017)Zeyer, Doetsch, Voigtlaender, Schl{\"u}ter, and
  Ney}]{zeyer2017comprehensive}
\bibinfo{author}{A.~Zeyer}, \bibinfo{author}{P.~Doetsch},
  \bibinfo{author}{P.~Voigtlaender}, \bibinfo{author}{R.~Schl{\"u}ter},
  \bibinfo{author}{H.~Ney},
\newblock \bibinfo{title}{A comprehensive study of deep bidirectional {LSTM}
  {RNNs} for acoustic modeling in speech recognition},
\newblock in: \bibinfo{booktitle}{2017 IEEE international conference on
  acoustics, speech and signal processing (ICASSP)},
  \bibinfo{organization}{IEEE}, \bibinfo{year}{2017}, pp.
  \bibinfo{pages}{2462--2466}.
%Type = Article
\bibitem[{Liu and Guo(2019)}]{liu2019bidirectional}
\bibinfo{author}{G.~Liu}, \bibinfo{author}{J.~Guo},
\newblock \bibinfo{title}{Bidirectional {LSTM} with attention mechanism and
  convolutional layer for text classification},
\newblock \bibinfo{journal}{Neurocomputing} \bibinfo{volume}{337}
  (\bibinfo{year}{2019}) \bibinfo{pages}{325--338}.
%Type = Article
\bibitem[{Liu et~al.(2016)Liu, Springer, Li, Moody, Juan, Chorro, Castells,
  Roig, Silva, Johnson et~al.}]{liu2016open}
\bibinfo{author}{C.~Liu}, \bibinfo{author}{D.~Springer},
  \bibinfo{author}{Q.~Li}, \bibinfo{author}{B.~Moody}, \bibinfo{author}{R.~A.
  Juan}, \bibinfo{author}{F.~J. Chorro}, \bibinfo{author}{F.~Castells},
  \bibinfo{author}{J.~M. Roig}, \bibinfo{author}{I.~Silva},
  \bibinfo{author}{A.~E. Johnson}, et~al.,
\newblock \bibinfo{title}{An open access database for the evaluation of heart
  sound algorithms},
\newblock \bibinfo{journal}{Physiological Measurement} \bibinfo{volume}{37}
  (\bibinfo{year}{2016}) \bibinfo{pages}{2181}.
%Type = Inproceedings
\bibitem[{Huai et~al.(2020)Huai, Panote, Choi, and Kuwahara}]{huai2020heart}
\bibinfo{author}{X.~Huai}, \bibinfo{author}{S.~Panote},
  \bibinfo{author}{D.~Choi}, \bibinfo{author}{N.~Kuwahara},
\newblock \bibinfo{title}{Heart sound recognition technology based on deep
  learning},
\newblock in: \bibinfo{booktitle}{International Conference on Human-Computer
  Interaction}, \bibinfo{organization}{Springer}, \bibinfo{year}{2020}, pp.
  \bibinfo{pages}{491--500}.
%Type = Article
\bibitem[{Rubin et~al.(2017)Rubin, Abreu, Ganguli, Nelaturi, Matei, and
  Sricharan}]{rubin2017recognizing}
\bibinfo{author}{J.~Rubin}, \bibinfo{author}{R.~Abreu},
  \bibinfo{author}{A.~Ganguli}, \bibinfo{author}{S.~Nelaturi},
  \bibinfo{author}{I.~Matei}, \bibinfo{author}{K.~Sricharan},
\newblock \bibinfo{title}{Recognizing abnormal heart sounds using deep
  learning},
\newblock \bibinfo{journal}{arXiv preprint arXiv:1707.04642}
  (\bibinfo{year}{2017}).
%Type = Article
\bibitem[{Demir et~al.(2019)Demir, {\c{S}}eng{\"u}r, Bajaj, and
  Polat}]{demir2019towards}
\bibinfo{author}{F.~Demir}, \bibinfo{author}{A.~{\c{S}}eng{\"u}r},
  \bibinfo{author}{V.~Bajaj}, \bibinfo{author}{K.~Polat},
\newblock \bibinfo{title}{Towards the classification of heart sounds based on
  convolutional deep neural network},
\newblock \bibinfo{journal}{Health information science and systems}
  \bibinfo{volume}{7} (\bibinfo{year}{2019}) \bibinfo{pages}{1--9}.
%Type = Article
\bibitem[{Dominguez-Morales et~al.(2017)Dominguez-Morales, Jimenez-Fernandez,
  Dominguez-Morales, and Jimenez-Moreno}]{dominguez2017deep}
\bibinfo{author}{J.~P. Dominguez-Morales}, \bibinfo{author}{A.~F.
  Jimenez-Fernandez}, \bibinfo{author}{M.~J. Dominguez-Morales},
  \bibinfo{author}{G.~Jimenez-Moreno},
\newblock \bibinfo{title}{Deep neural networks for the recognition and
  classification of heart murmurs using neuromorphic auditory sensors},
\newblock \bibinfo{journal}{IEEE transactions on biomedical circuits and
  systems} \bibinfo{volume}{12} (\bibinfo{year}{2017}) \bibinfo{pages}{24--34}.
%Type = Article
\bibitem[{Cheng et~al.(2019)Cheng, Huang, Li, and Gui}]{cheng2019design}
\bibinfo{author}{X.~Cheng}, \bibinfo{author}{J.~Huang},
  \bibinfo{author}{Y.~Li}, \bibinfo{author}{G.~Gui},
\newblock \bibinfo{title}{Design and application of a laconic heart sound
  neural network},
\newblock \bibinfo{journal}{IEEE Access} \bibinfo{volume}{7}
  (\bibinfo{year}{2019}) \bibinfo{pages}{124417--124425}.
%Type = Article
\bibitem[{Ergen et~al.(2012)Ergen, Tatar, and Gulcur}]{ergen2012time}
\bibinfo{author}{B.~Ergen}, \bibinfo{author}{Y.~Tatar}, \bibinfo{author}{H.~O.
  Gulcur},
\newblock \bibinfo{title}{Time-frequency analysis of phonocardiogram signals
  using wavelet transform: a comparative study},
\newblock \bibinfo{journal}{Computer methods in biomechanics and biomedical
  engineering} \bibinfo{volume}{15} (\bibinfo{year}{2012})
  \bibinfo{pages}{371--381}.
%Type = Book
\bibitem[{Stankovic et~al.(2014)Stankovic, Dakovi{\'c}, and
  Thayaparan}]{stankovic2014time}
\bibinfo{author}{L.~Stankovic}, \bibinfo{author}{M.~Dakovi{\'c}},
  \bibinfo{author}{T.~Thayaparan}, \bibinfo{title}{Time-frequency signal
  analysis with applications}, \bibinfo{publisher}{Artech house},
  \bibinfo{year}{2014}.
%Type = Article
\bibitem[{Mahmoud et~al.(2006)Mahmoud, Hussain, Cosic, and
  Fang}]{mahmoud2006time}
\bibinfo{author}{S.~S. Mahmoud}, \bibinfo{author}{Z.~M. Hussain},
  \bibinfo{author}{I.~Cosic}, \bibinfo{author}{Q.~Fang},
\newblock \bibinfo{title}{Time-frequency analysis of normal and abnormal
  biological signals},
\newblock \bibinfo{journal}{Biomedical Signal Processing and Control}
  \bibinfo{volume}{1} (\bibinfo{year}{2006}) \bibinfo{pages}{33--43}.
%Type = Article
\bibitem[{Obaidat(1993)}]{obaidat1993phonocardiogram}
\bibinfo{author}{M.~Obaidat},
\newblock \bibinfo{title}{Phonocardiogram signal analysis: techniques and
  performance comparison},
\newblock \bibinfo{journal}{Journal of medical engineering \& technology}
  \bibinfo{volume}{17} (\bibinfo{year}{1993}) \bibinfo{pages}{221--227}.
%Type = Article
\bibitem[{Peng et~al.(2011)Peng, Meng, Chu, Lang, Zhang, and
  Yang}]{peng2011polynomial}
\bibinfo{author}{Z.~Peng}, \bibinfo{author}{G.~Meng}, \bibinfo{author}{F.~Chu},
  \bibinfo{author}{Z.~Lang}, \bibinfo{author}{W.~Zhang},
  \bibinfo{author}{Y.~Yang},
\newblock \bibinfo{title}{Polynomial chirplet transform with application to
  instantaneous frequency estimation},
\newblock \bibinfo{journal}{IEEE Transactions on Instrumentation and
  Measurement} \bibinfo{volume}{60} (\bibinfo{year}{2011})
  \bibinfo{pages}{3222--3229}.
%Type = Article
\bibitem[{Auger et~al.(1996)Auger, Flandrin, Gon{\c{c}}alv{\`e}s, and
  Lemoine}]{auger1996time}
\bibinfo{author}{F.~Auger}, \bibinfo{author}{P.~Flandrin},
  \bibinfo{author}{P.~Gon{\c{c}}alv{\`e}s}, \bibinfo{author}{O.~Lemoine},
\newblock \bibinfo{title}{Time-frequency toolbox},
\newblock \bibinfo{journal}{CNRS France-Rice University} \bibinfo{volume}{46}
  (\bibinfo{year}{1996}).
%Type = Article
\bibitem[{Cherif et~al.(2010)Cherif, Debbal, and
  Bereksi-Reguig}]{cherif2010choice}
\bibinfo{author}{L.~H. Cherif}, \bibinfo{author}{S.~Debbal},
  \bibinfo{author}{F.~Bereksi-Reguig},
\newblock \bibinfo{title}{Choice of the wavelet analyzing in the
  phonocardiogram signal analysis using the discrete and the packet wavelet
  transform},
\newblock \bibinfo{journal}{Expert Systems with Applications}
  \bibinfo{volume}{37} (\bibinfo{year}{2010}) \bibinfo{pages}{913--918}.
%Type = Inproceedings
\bibitem[{Vikhe et~al.(2009)Vikhe, Hamde, and Nehe}]{vikhe2009wavelet}
\bibinfo{author}{P.~Vikhe}, \bibinfo{author}{S.~Hamde},
  \bibinfo{author}{N.~Nehe},
\newblock \bibinfo{title}{Wavelet transform based abnormality analysis of heart
  sound},
\newblock in: \bibinfo{booktitle}{2009 International Conference on Advances in
  Computing, Control, and Telecommunication Technologies},
  \bibinfo{organization}{IEEE}, \bibinfo{year}{2009}, pp.
  \bibinfo{pages}{367--371}.
%Type = Article
\bibitem[{Debbal and Bereksi-Reguig(2004)}]{debbal2004analysis}
\bibinfo{author}{S.~Debbal}, \bibinfo{author}{F.~Bereksi-Reguig},
\newblock \bibinfo{title}{Analysis of the second heart sound using continuous
  wavelet transform},
\newblock \bibinfo{journal}{Journal of medical engineering \& technology}
  \bibinfo{volume}{28} (\bibinfo{year}{2004}) \bibinfo{pages}{151--156}.
%Type = Inproceedings
\bibitem[{Taebi and Mansy(2017)}]{taebi2017analysis}
\bibinfo{author}{A.~Taebi}, \bibinfo{author}{H.~A. Mansy},
\newblock \bibinfo{title}{Analysis of seismocardiographic signals using
  polynomial chirplet transform and smoothed pseudo {Wigner-Ville}
  distribution},
\newblock in: \bibinfo{booktitle}{2017 IEEE Signal Processing in Medicine and
  Biology Symposium (SPMB)}, \bibinfo{organization}{IEEE},
  \bibinfo{year}{2017}, pp. \bibinfo{pages}{1--6}.
%Type = Article
\bibitem[{Mann and Haykin(1995)}]{mann1995chirplet}
\bibinfo{author}{S.~Mann}, \bibinfo{author}{S.~Haykin},
\newblock \bibinfo{title}{The chirplet transform: Physical considerations},
\newblock \bibinfo{journal}{IEEE Transactions on Signal Processing}
  \bibinfo{volume}{43} (\bibinfo{year}{1995}) \bibinfo{pages}{2745--2761}.
%Type = Article
\bibitem[{Ghosh et~al.(2020)Ghosh, Ponnalagu, Tripathy, and
  Acharya}]{ghosh2020automated}
\bibinfo{author}{S.~K. Ghosh}, \bibinfo{author}{R.~Ponnalagu},
  \bibinfo{author}{R.~Tripathy}, \bibinfo{author}{U.~R. Acharya},
\newblock \bibinfo{title}{Automated detection of heart valve diseases using
  chirplet transform and multiclass composite classifier with {PCG} signals},
\newblock \bibinfo{journal}{Computers in biology and medicine}
  \bibinfo{volume}{118} (\bibinfo{year}{2020}) \bibinfo{pages}{103632}.
%Type = Article
\bibitem[{Djebbari and Bereksi-Reguig(2013)}]{djebbari2013detection}
\bibinfo{author}{A.~Djebbari}, \bibinfo{author}{F.~Bereksi-Reguig},
\newblock \bibinfo{title}{Detection of the valvular split within the second
  heart sound using the reassigned smoothed pseudo {Wigner-Ville}
  distribution},
\newblock \bibinfo{journal}{Biomedical engineering online} \bibinfo{volume}{12}
  (\bibinfo{year}{2013}) \bibinfo{pages}{1--21}.
%Type = Article
\bibitem[{Chen et~al.(2015)Chen, Xiang, and Zhang}]{chen2015recognition}
\bibinfo{author}{T.~Chen}, \bibinfo{author}{L.~Xiang},
  \bibinfo{author}{M.~Zhang},
\newblock \bibinfo{title}{Recognition of heart sound based on distribution of
  {Choi-Williams}},
\newblock \bibinfo{journal}{Research on Biomedical Engineering}
  \bibinfo{volume}{31} (\bibinfo{year}{2015}) \bibinfo{pages}{189--195}.
%Type = Article
\bibitem[{Taebi and Mansy(2017)}]{taebi2017time}
\bibinfo{author}{A.~Taebi}, \bibinfo{author}{H.~A. Mansy},
\newblock \bibinfo{title}{Time-frequency distribution of seismocardiographic
  signals: A comparative study},
\newblock \bibinfo{journal}{Bioengineering} \bibinfo{volume}{4}
  (\bibinfo{year}{2017}) \bibinfo{pages}{32}.
%Type = Article
\bibitem[{Bozkurt et~al.(2018)Bozkurt, Germanakis, and
  Stylianou}]{bozkurt2018study}
\bibinfo{author}{B.~Bozkurt}, \bibinfo{author}{I.~Germanakis},
  \bibinfo{author}{Y.~Stylianou},
\newblock \bibinfo{title}{A study of time-frequency features for {CNN}-based
  automatic heart sound classification for pathology detection},
\newblock \bibinfo{journal}{Computers in biology and medicine}
  \bibinfo{volume}{100} (\bibinfo{year}{2018}) \bibinfo{pages}{132--143}.
%Type = Inproceedings
\bibitem[{Ren et~al.(2018)Ren, Cummins, Pandit, Han, Qian, and
  Schuller}]{ren2018learning}
\bibinfo{author}{Z.~Ren}, \bibinfo{author}{N.~Cummins},
  \bibinfo{author}{V.~Pandit}, \bibinfo{author}{J.~Han},
  \bibinfo{author}{K.~Qian}, \bibinfo{author}{B.~Schuller},
\newblock \bibinfo{title}{Learning image-based representations for heart sound
  classification},
\newblock in: \bibinfo{booktitle}{Proceedings of the 2018 International
  Conference on Digital Health}, \bibinfo{year}{2018}, pp.
  \bibinfo{pages}{143--147}.
%Type = Inproceedings
\bibitem[{Nilanon et~al.(2016)Nilanon, Yao, Hao, Purushotham, and
  Liu}]{nilanon2016normal}
\bibinfo{author}{T.~Nilanon}, \bibinfo{author}{J.~Yao},
  \bibinfo{author}{J.~Hao}, \bibinfo{author}{S.~Purushotham},
  \bibinfo{author}{Y.~Liu},
\newblock \bibinfo{title}{Normal/abnormal heart sound recordings classification
  using convolutional neural network},
\newblock in: \bibinfo{booktitle}{2016 computing in cardiology conference
  (CinC)}, \bibinfo{organization}{IEEE}, \bibinfo{year}{2016}, pp.
  \bibinfo{pages}{585--588}.
%Type = Article
\bibitem[{Rizal et~al.(2020)Rizal, Adz-Dzikri, Arik, and
  Fauzi}]{rizalclassification}
\bibinfo{author}{A.~Rizal}, \bibinfo{author}{A.~Adz-Dzikri},
  \bibinfo{author}{M.~Arik}, \bibinfo{author}{G.~Fauzi},
\newblock \bibinfo{title}{Classification of normal and abnormal heart sound
  using continuous wavelet transform and {ResNet-50}},
\newblock \bibinfo{journal}{Technology Reports of Kansai University}
  \bibinfo{volume}{62} (\bibinfo{year}{2020}) \bibinfo{pages}{2595--2601}.
%Type = Inproceedings
\bibitem[{Alaskar et~al.(2019)Alaskar, Alzhrani, Hussain, and
  Almarshed}]{alaskar2019implementation}
\bibinfo{author}{H.~Alaskar}, \bibinfo{author}{N.~Alzhrani},
  \bibinfo{author}{A.~Hussain}, \bibinfo{author}{F.~Almarshed},
\newblock \bibinfo{title}{The implementation of pretrained {AlexNet} on {PCG}
  classification},
\newblock in: \bibinfo{booktitle}{International Conference on Intelligent
  Computing}, \bibinfo{organization}{Springer}, \bibinfo{year}{2019}, pp.
  \bibinfo{pages}{784--794}.
%Type = Article
\bibitem[{Jung et~al.(2021)Jung, Liao, Wu, Yuan, and Sun}]{jung2021efficiently}
\bibinfo{author}{S.-Y. Jung}, \bibinfo{author}{C.-H. Liao},
  \bibinfo{author}{Y.-S. Wu}, \bibinfo{author}{S.-M. Yuan},
  \bibinfo{author}{C.-T. Sun},
\newblock \bibinfo{title}{Efficiently classifying lung sounds through depthwise
  separable {CNN} models with fused {STFT} and {MFCC} features},
\newblock \bibinfo{journal}{Diagnostics} \bibinfo{volume}{11}
  (\bibinfo{year}{2021}) \bibinfo{pages}{732}.
%Type = Article
\bibitem[{Zhang et~al.(2019)Zhang, Wang, Bao, Wang, and Xu}]{zhang2019large}
\bibinfo{author}{L.~Zhang}, \bibinfo{author}{D.~Wang},
  \bibinfo{author}{C.~Bao}, \bibinfo{author}{Y.~Wang}, \bibinfo{author}{K.~Xu},
\newblock \bibinfo{title}{Large-scale whale-call classification by transfer
  learning on multi-scale waveforms and time-frequency features},
\newblock \bibinfo{journal}{Applied Sciences} \bibinfo{volume}{9}
  (\bibinfo{year}{2019}) \bibinfo{pages}{1020}.
%Type = Article
\bibitem[{McLoughlin et~al.(2020)McLoughlin, Xie, Song, Phan, and
  Palaniappan}]{mcloughlin2020time}
\bibinfo{author}{I.~McLoughlin}, \bibinfo{author}{Z.~Xie},
  \bibinfo{author}{Y.~Song}, \bibinfo{author}{H.~Phan},
  \bibinfo{author}{R.~Palaniappan},
\newblock \bibinfo{title}{Time-frequency feature fusion for noise robust audio
  event classification},
\newblock \bibinfo{journal}{Circuits, Systems, and Signal Processing}
  \bibinfo{volume}{39} (\bibinfo{year}{2020}) \bibinfo{pages}{1672--1687}.
%Type = Article
\bibitem[{Jalayer et~al.(2021)Jalayer, Orsenigo, and
  Vercellis}]{jalayer2021fault}
\bibinfo{author}{M.~Jalayer}, \bibinfo{author}{C.~Orsenigo},
  \bibinfo{author}{C.~Vercellis},
\newblock \bibinfo{title}{Fault detection and diagnosis for rotating machinery:
  A model based on convolutional {LSTM}, fast fourier and continuous wavelet
  transforms},
\newblock \bibinfo{journal}{Computers in Industry} \bibinfo{volume}{125}
  (\bibinfo{year}{2021}) \bibinfo{pages}{103378}.
%Type = Article
\bibitem[{Liu et~al.(2021{\natexlab{a}})Liu, Lv, Shen, Xiong, Yang, and
  Liu}]{liu2021multiscale}
\bibinfo{author}{X.~Liu}, \bibinfo{author}{L.~Lv}, \bibinfo{author}{Y.~Shen},
  \bibinfo{author}{P.~Xiong}, \bibinfo{author}{J.~Yang},
  \bibinfo{author}{J.~Liu},
\newblock \bibinfo{title}{Multiscale space-time-frequency feature-guided
  multitask learning {CNN} for motor imagery {EEG} classification},
\newblock \bibinfo{journal}{Journal of Neural Engineering} \bibinfo{volume}{18}
  (\bibinfo{year}{2021}{\natexlab{a}}) \bibinfo{pages}{026003}.
%Type = Article
\bibitem[{Liu et~al.(2021{\natexlab{b}})Liu, Han, Tian, Zhou, and
  Liu}]{liu2021ecg}
\bibinfo{author}{G.~Liu}, \bibinfo{author}{X.~Han}, \bibinfo{author}{L.~Tian},
  \bibinfo{author}{W.~Zhou}, \bibinfo{author}{H.~Liu},
\newblock \bibinfo{title}{{ECG} quality assessment based on hand-crafted
  statistics and deep-learned {S-transform} spectrogram features},
\newblock \bibinfo{journal}{Computer Methods and Programs in Biomedicine}
  \bibinfo{volume}{208} (\bibinfo{year}{2021}{\natexlab{b}})
  \bibinfo{pages}{106269}.
%Type = Article
\bibitem[{Yan and Jia(2018)}]{yan2018novel}
\bibinfo{author}{X.~Yan}, \bibinfo{author}{M.~Jia},
\newblock \bibinfo{title}{A novel optimized {SVM} classification algorithm with
  multi-domain feature and its application to fault diagnosis of rolling
  bearing},
\newblock \bibinfo{journal}{Neurocomputing} \bibinfo{volume}{313}
  (\bibinfo{year}{2018}) \bibinfo{pages}{47--64}.
%Type = Article
\bibitem[{Cui and Wang(2017)}]{cui2017biosignal}
\bibinfo{author}{J.~Cui}, \bibinfo{author}{D.~Wang},
\newblock \bibinfo{title}{Biosignal analysis with matching-pursuit based
  adaptive chirplet transform},
\newblock \bibinfo{journal}{arXiv preprint arXiv:1709.08328}
  (\bibinfo{year}{2017}).
%Type = Article
\bibitem[{Russakovsky et~al.(2015)Russakovsky, Deng, Su, Krause, Satheesh, Ma,
  Huang, Karpathy, Khosla, Bernstein et~al.}]{russakovsky2015imagenet}
\bibinfo{author}{O.~Russakovsky}, \bibinfo{author}{J.~Deng},
  \bibinfo{author}{H.~Su}, \bibinfo{author}{J.~Krause},
  \bibinfo{author}{S.~Satheesh}, \bibinfo{author}{S.~Ma},
  \bibinfo{author}{Z.~Huang}, \bibinfo{author}{A.~Karpathy},
  \bibinfo{author}{A.~Khosla}, \bibinfo{author}{M.~Bernstein}, et~al.,
\newblock \bibinfo{title}{Imagenet large scale visual recognition challenge},
\newblock \bibinfo{journal}{International journal of computer vision}
  \bibinfo{volume}{115} (\bibinfo{year}{2015}) \bibinfo{pages}{211--252}.
%Type = Inproceedings
\bibitem[{Howard et~al.(2019)Howard, Sandler, Chu, Chen, Chen, Tan, Wang, Zhu,
  Pang, Vasudevan et~al.}]{howard2019searching}
\bibinfo{author}{A.~Howard}, \bibinfo{author}{M.~Sandler},
  \bibinfo{author}{G.~Chu}, \bibinfo{author}{L.-C. Chen},
  \bibinfo{author}{B.~Chen}, \bibinfo{author}{M.~Tan},
  \bibinfo{author}{W.~Wang}, \bibinfo{author}{Y.~Zhu},
  \bibinfo{author}{R.~Pang}, \bibinfo{author}{V.~Vasudevan}, et~al.,
\newblock \bibinfo{title}{Searching for {MobilenetV3}},
\newblock in: \bibinfo{booktitle}{Proceedings of the IEEE/CVF International
  Conference on Computer Vision}, \bibinfo{year}{2019}, pp.
  \bibinfo{pages}{1314--1324}.
%Type = Inproceedings
\bibitem[{He et~al.(2016)He, Zhang, Ren, and Sun}]{he2016deep}
\bibinfo{author}{K.~He}, \bibinfo{author}{X.~Zhang}, \bibinfo{author}{S.~Ren},
  \bibinfo{author}{J.~Sun},
\newblock \bibinfo{title}{Deep residual learning for image recognition},
\newblock in: \bibinfo{booktitle}{Proceedings of the IEEE conference on
  computer vision and pattern recognition}, \bibinfo{year}{2016}, pp.
  \bibinfo{pages}{770--778}.
%Type = Inproceedings
\bibitem[{Hu et~al.(2018)Hu, Shen, and Sun}]{hu2018squeeze}
\bibinfo{author}{J.~Hu}, \bibinfo{author}{L.~Shen}, \bibinfo{author}{G.~Sun},
\newblock \bibinfo{title}{Squeeze-and-excitation networks},
\newblock in: \bibinfo{booktitle}{Proceedings of the IEEE conference on
  computer vision and pattern recognition}, \bibinfo{year}{2018}, pp.
  \bibinfo{pages}{7132--7141}.
%Type = Inproceedings
\bibitem[{Huang et~al.(2017)Huang, Liu, Van Der~Maaten, and
  Weinberger}]{huang2017densely}
\bibinfo{author}{G.~Huang}, \bibinfo{author}{Z.~Liu}, \bibinfo{author}{L.~Van
  Der~Maaten}, \bibinfo{author}{K.~Q. Weinberger},
\newblock \bibinfo{title}{Densely connected convolutional networks},
\newblock in: \bibinfo{booktitle}{Proceedings of the IEEE conference on
  computer vision and pattern recognition}, \bibinfo{year}{2017}, pp.
  \bibinfo{pages}{4700--4708}.
%Type = Inproceedings
\bibitem[{Tan et~al.(2019)Tan, Chen, Pang, Vasudevan, Sandler, Howard, and
  Le}]{tan2019mnasnet}
\bibinfo{author}{M.~Tan}, \bibinfo{author}{B.~Chen}, \bibinfo{author}{R.~Pang},
  \bibinfo{author}{V.~Vasudevan}, \bibinfo{author}{M.~Sandler},
  \bibinfo{author}{A.~Howard}, \bibinfo{author}{Q.~V. Le},
\newblock \bibinfo{title}{{MNasNet}: Platform-aware neural architecture search
  for mobile},
\newblock in: \bibinfo{booktitle}{Proceedings of the IEEE/CVF Conference on
  Computer Vision and Pattern Recognition}, \bibinfo{year}{2019}, pp.
  \bibinfo{pages}{2820--2828}.
%Type = Inproceedings
\bibitem[{Yang et~al.(2018)Yang, Howard, Chen, Zhang, Go, Sandler, Sze, and
  Adam}]{yang2018netadapt}
\bibinfo{author}{T.-J. Yang}, \bibinfo{author}{A.~Howard},
  \bibinfo{author}{B.~Chen}, \bibinfo{author}{X.~Zhang},
  \bibinfo{author}{A.~Go}, \bibinfo{author}{M.~Sandler},
  \bibinfo{author}{V.~Sze}, \bibinfo{author}{H.~Adam},
\newblock \bibinfo{title}{Netadapt: Platform-aware neural network adaptation
  for mobile applications},
\newblock in: \bibinfo{booktitle}{Proceedings of the European Conference on
  Computer Vision (ECCV)}, \bibinfo{year}{2018}, pp. \bibinfo{pages}{285--300}.
%Type = Article
\bibitem[{Bao et~al.(2022)Bao, Xu, and Kamavuako}]{bao2022effect}
\bibinfo{author}{X.~Bao}, \bibinfo{author}{Y.~Xu}, \bibinfo{author}{E.~N.
  Kamavuako},
\newblock \bibinfo{title}{The effect of signal duration on the classification
  of heart sounds: A deep learning approach},
\newblock \bibinfo{journal}{Sensors} \bibinfo{volume}{22}
  (\bibinfo{year}{2022}) \bibinfo{pages}{2261}.
%Type = Article
\bibitem[{Paszke et~al.(2019)Paszke, Gross, Massa, Lerer, Bradbury, Chanan,
  Killeen, Lin, Gimelshein, Antiga et~al.}]{paszke2019pytorch}
\bibinfo{author}{A.~Paszke}, \bibinfo{author}{S.~Gross},
  \bibinfo{author}{F.~Massa}, \bibinfo{author}{A.~Lerer},
  \bibinfo{author}{J.~Bradbury}, \bibinfo{author}{G.~Chanan},
  \bibinfo{author}{T.~Killeen}, \bibinfo{author}{Z.~Lin},
  \bibinfo{author}{N.~Gimelshein}, \bibinfo{author}{L.~Antiga}, et~al.,
\newblock \bibinfo{title}{Pytorch: An imperative style, high-performance deep
  learning library},
\newblock \bibinfo{journal}{Advances in neural information processing systems}
  \bibinfo{volume}{32} (\bibinfo{year}{2019}).
%Type = Misc
\bibitem[{Wightman(2019)}]{rw2019timm}
\bibinfo{author}{R.~Wightman}, \bibinfo{title}{Pytorch image models},
  \bibinfo{howpublished}{\url{https://github.com/rwightman/pytorch-image-models}},
  \bibinfo{year}{2019}. \DOIprefix\doi{10.5281/zenodo.4414861}.

\end{thebibliography}

% Biography
\subsection*{}
\setlength\intextsep{0pt} % align top of photo with text
\begin{wrapfigure}{l}{0.13\textwidth}
    \centering
    \includegraphics[width=0.15\textwidth]{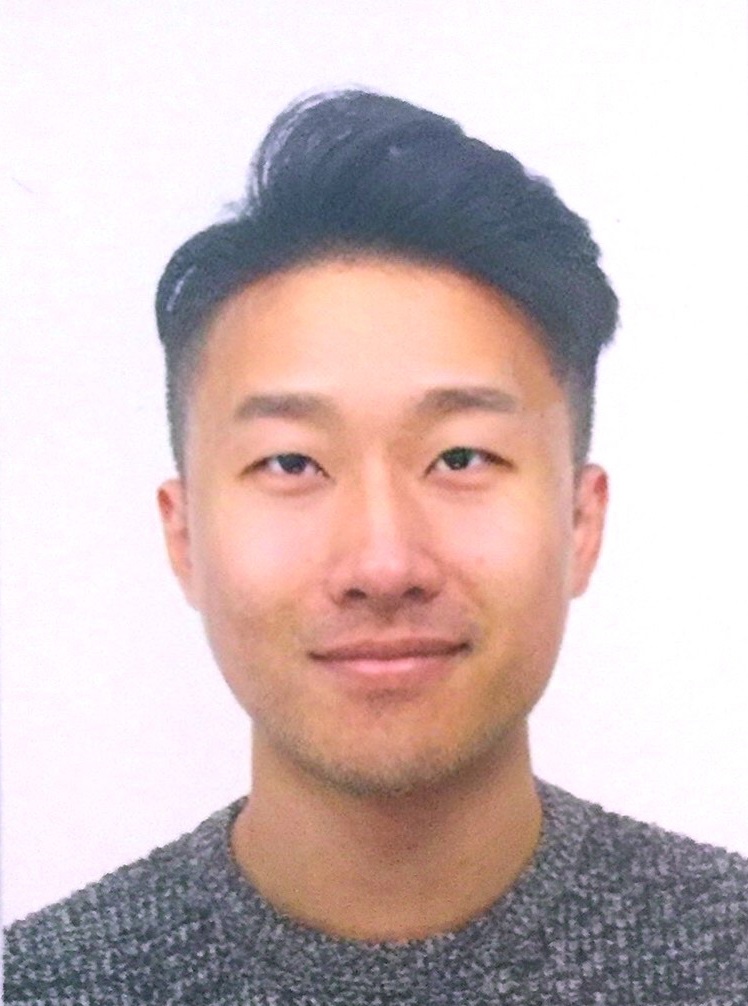}
\end{wrapfigure}

\noindent \textbf{Xinqi Bao} is currently a PhD student in the Centre for Robotics Research (CoRE) at Department of Engineering, King's College London. His research focuses on building low-cost and portable device to capture physiological signals, and designing machine learning algorithms to conduct automated diagnosis.

\subsection*{}
\setlength\intextsep{0pt} % align top of photo with text
\begin{wrapfigure}{l}{0.13\textwidth}
    \centering
    \includegraphics[width=0.15\textwidth]{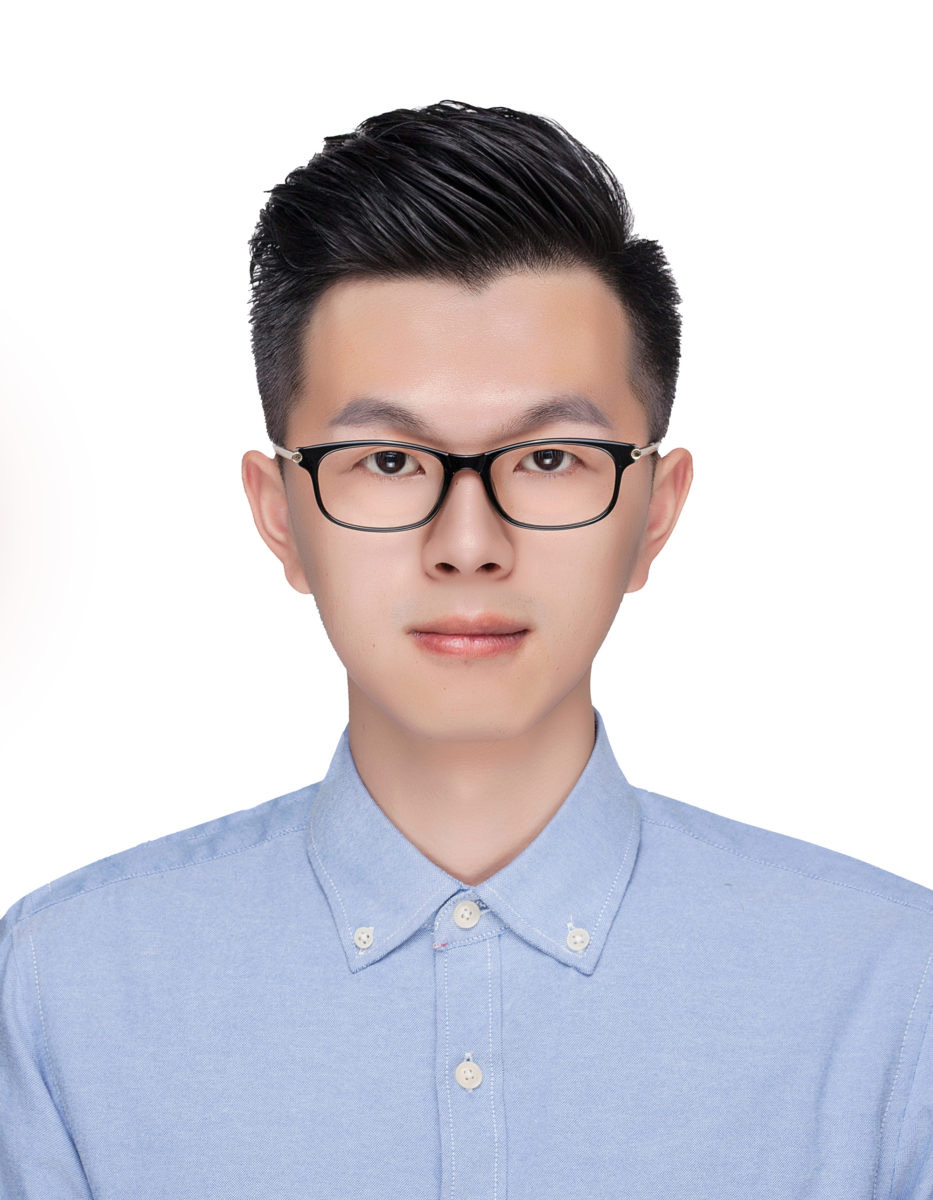}
\end{wrapfigure}

\noindent \textbf{Yujia Xu} is a PhD student in the Centre for Robotics Research at King’s College London. He is interested in deep learning, medical imaging, signal processing, and generative adversarial networks. His current research focuses on developing reliable automatic computer-aided classification systems using deep learning.

\subsection*{}
\setlength\intextsep{0pt} % align top of photo with text
\begin{wrapfigure}{l}{0.13\textwidth}
    \centering
    \includegraphics[width=0.15\textwidth]{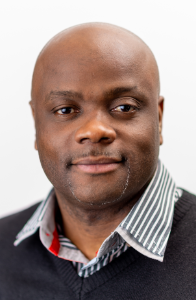}
\end{wrapfigure}

\noindent \textbf{Ernest N. Kamavuako} joined King’s College London as Senior Lecturer in the department of Informatics in October 2017. Between 2014 and 2017, he held a position as Associate Professor in the Department of Health Science and Technology, Aalborg University, Denmark with excellent teaching and supervision skills. From 2007 to 2008, he was a Research Scholar in the Biomedical Department, IUPUI, Indianapolis, USA. From 2012 to 2013, he was a Visiting Postdoctoral Fellow at the Institute of Biomedical Engineering, University of New Brunswick, Canada. Since January 2017, ENK is appointed Adjunct Professor in the Department of electrical and computer engineering at the University of New Brunswick. In 2017 he was an academic visitor in the department of Bioengineering, Imperial College London, United Kingdom.  He has good publication record with main research interests related to the use of invasive recordings in the control of upper limb prostheses. Other research interests include neurorehabilitation, applied signal processing and Health technologies to promote wellbeing.  

\subsection*{}
\setlength\intextsep{0pt}
\begin{wrapfigure}{l}{0.13\textwidth}
    \centering
    \includegraphics[width=0.15\textwidth]{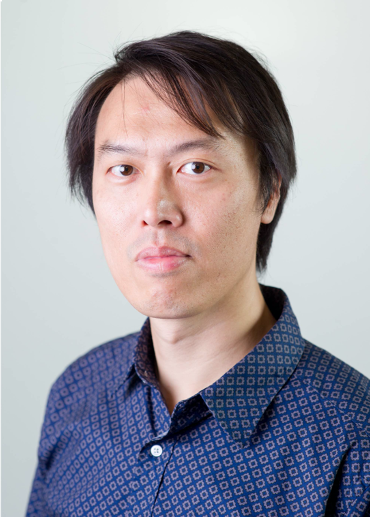}
\end{wrapfigure}
\noindent \textbf{Hak-Keung Lam} (Fellow, IEEE) Hak-Keung Lam (Fellow, IEEE) received the B.Eng. (Hons.) and Ph.D. degrees from the Department of Electronic and Information Engineering, The Hong Kong Polytechnic University, Hong Kong, China, in 1995 and 2000, respectively. During the period of 2000 and 2005, he worked with the Department of Electronic and Information Engineering at The Hong Kong Polytechnic University as Post- Doctoral Fellow and Research Fellow, respectively. He joined as a Lecturer at King’s College London in 2005 and is currently a Reader.

His current research interests include intelligent control systems and computational intelligence. He has served as a program committee member and international advisory board member for various international conferences and a reviewer for various books, international journals and international conferences. He is an Associate Editor for IEEE Transactions on Fuzzy Systems, IEEE Transactions on Circuits and Systems II: Express Briefs, IET Control Theory and Applications, International Journal of Fuzzy Systems, Neurocomputing, and Nonlinear Dynamics; and Guest Editor for a number of international journals, and is in the editorial board of for a number of international journals. He was named as Highly Cited Researcher and is an IEEE Fellow. 

He is the coeditor for two edited volumes: Control of Chaotic Nonlinear Circuits (World Scientific, 2009) and Computational Intelligence and Its Applications (World Scientific, 2012), and the coauthor of the monographs: Stability Analysis of Fuzzy-Model-Based Control Systems (Springer, 2011), Polynomial Fuzzy Model Based Control Systems (Springer, 2016), and Analysis and Synthesis for Interval Type-2 Fuzzy-Model-Based Systems (Springer, 2016). 

\subsection*{}
\setlength\intextsep{0pt} % align top of photo with text
\begin{wrapfigure}{l}{0.13\textwidth}
    \centering
    \includegraphics[width=0.15\textwidth]{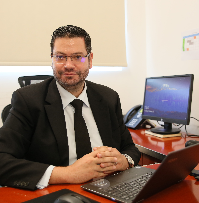}
\end{wrapfigure}

\noindent \textbf{Mohamed Trabelsi} (Senior Member, IEEE) received the B.Sc. degree in electrical engineering from INSAT, Tunisia, in 2006, and the M.Sc. degree in automated systems and the Ph.D. degree in energy systems from INSA Lyon, France, in 2006 and 2009, respectively.,From October 2009 to August 2018, he held different research positions at Qatar University and Texas A\&M University at Qatar. Since September 2018, he has been an Associate Professor with the Kuwait College of Science and Technology. He has published more than 120 journals and conference papers. He is the author of two books and two book chapters. His research interests include systems control with applications arising in the contexts of power electronics, energy conversion, renewable energies integration, and smart grids.

\subsection*{}
\setlength\intextsep{0pt} % align top of photo with text
\begin{wrapfigure}{l}{0.13\textwidth}
    \centering
    \includegraphics[width=0.15\textwidth]{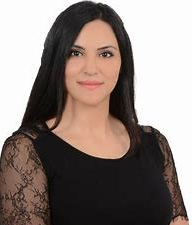}
\end{wrapfigure}

\noindent \textbf{Ines Chihi} Ines Chihi received the Ph.D. and Habilitation degrees from the National Engineering School of Tunis (ENIT), Tunisia, in 2013 and 2019, respectively. She was an Associate Professor of automation and industrial computing with the National Engineering School of Bizerte (ENIB), Tunisia, and a member of the Laboratory of Energy Applications and Renewable Energy Efficiency (LAPER), University Tunis El Manar. Now, she is an Assistant professor in Electrical Measurement and Sensor Technology,  Université du Luxembourg. Her research interests include intelligent modeling, control, and fault detection of complex systems with unpredictable behaviors.

\subsection*{}
\setlength\intextsep{0pt} % align top of photo with text
\begin{wrapfigure}{l}{0.13\textwidth}
    \centering
    \includegraphics[width=0.15\textwidth]{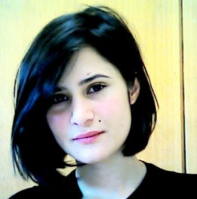}
\end{wrapfigure}

\noindent \textbf{Lilia Sidhom} received the Ph.D. degree in automation and industrial computing from the National Institute of Applied Sciences of Lyon (INSA Lyon), Villeurbanne, France, in 2011, and the University Habilitation degree from INSAT, Tunisia, in 2019. She is currently an Associate Professor on Automation and Industrial Computing with the Department of Mechanical Engineering, National Engineering School of Bizerta, Tunisia (ENIB), and a member of the Laboratory of Energy Applications and Renewable Energy Efficiency (LAPER), University Tunis El Manar. She received the automation and industrial computing engineering degree from the National Institute of Applied Sciences and Technology (INSAT), Tunis, Tunisia, in 2007. Her current research interests include advanced control of nonlinear complex and innovative system, soft sensors, smart diagnostic, observer/estimator. Her research applications are concerning smart grid, biomedical, drill string systems, and robots.

\end{document}